\documentclass[a4paper,twoside]{article}
\usepackage[sort]{natbib}
\bibpunct{(}{)}{,}{a}{}{,}
\usepackage[latin1]{inputenc}
\usepackage[english]{babel}
\usepackage{amsmath,amssymb,amsbsy}
\usepackage{graphicx}
\usepackage{rotating}
\usepackage{psfrag}
\usepackage{tabularx}
\usepackage{mparhack}
\usepackage[normalem]{ulem}
\newcommand\mptit\uline
\usepackage[hyphens]{url}
\usepackage[hidelinks,linktocpage]{hyperref}

\newcommand*\stabilityref[1]{
\medskip
{
\addtolength\leftskip{1ex}
\addtolength\rightskip{1ex}
\small
\noindent
#1
\par
}
}

\newcommand\eg{{e.g.}}
\newcommand\ie{{i.e.}}

\newcommand\rhs{{r.h.s.}}
\newcommand\mat[1]{\boldsymbol{\mathrm{#1}}}
\newcommand\nvec[1]{\boldsymbol{\mathrm{#1}}}
\newcommand\nodot{}
\renewcommand\d{\,\mathrm{d}}
\newcommand\de[2]{\frac{\partial #1}{\partial #2}}
\newcommand\deh[2]{{\partial #1}/{\partial #2}}
\newcommand\be{\begin{equation}}
\newcommand\ee[1]{\label{#1}\end{equation}}

\newcommand{\overbar}[1]{\mkern 3mu\overline{\mkern-3mu#1\mkern-3mu}\mkern 3mu}
\newcommand\Rey{\mbox{\textit{Re}}}
\newcommand\bnabla{\nabla}
\newcommand\op[1]{\mathcal{#1}}

\newcommand\trasp{\mathrm{T}}

\title{\sc An Introduction to Adjoint Problems}

\author{Paolo Luchini$^1$ and Alessandro Bottaro$^2$\\
$^1$DIIN, University of Salerno, Italy\\
\texttt{luchini@unisa.it}\\
$^2$DICCA, University of Genova, Italy\\
\texttt{alessandro.bottaro@unige.it}}

\date{}

\begin{document}
\maketitle

\let\subsubsection\subsection
\let\subsection\section
\makeatletter
\renewcommand*\l@section{\@dottedtocline{1}{0em}{1.5em}}
\makeatother

\noindent {\large Originally published as a Supplemental Appendix to\linebreak \href{https://doi.org/10.1146/annurev-fluid-010313-141253}{\textsc{Adjoint Equations in Stability Analysis}}, \textit{Annu. Rev. Fluid Mech.} \textbf{46}:493--517 (2014)}

\tableofcontents
\pagebreak

\subsection{Historical remarks}
\marginpar{
\vspace{-0.7cm}
\mptit{First use of adjoint equa\-tions} by Lagrange, founding member of the Turin Royal Society
\smallskip

\centering
\href{http://archive.org/details/mlangesdephilo03soci}{\includegraphics[width=\marginparwidth]{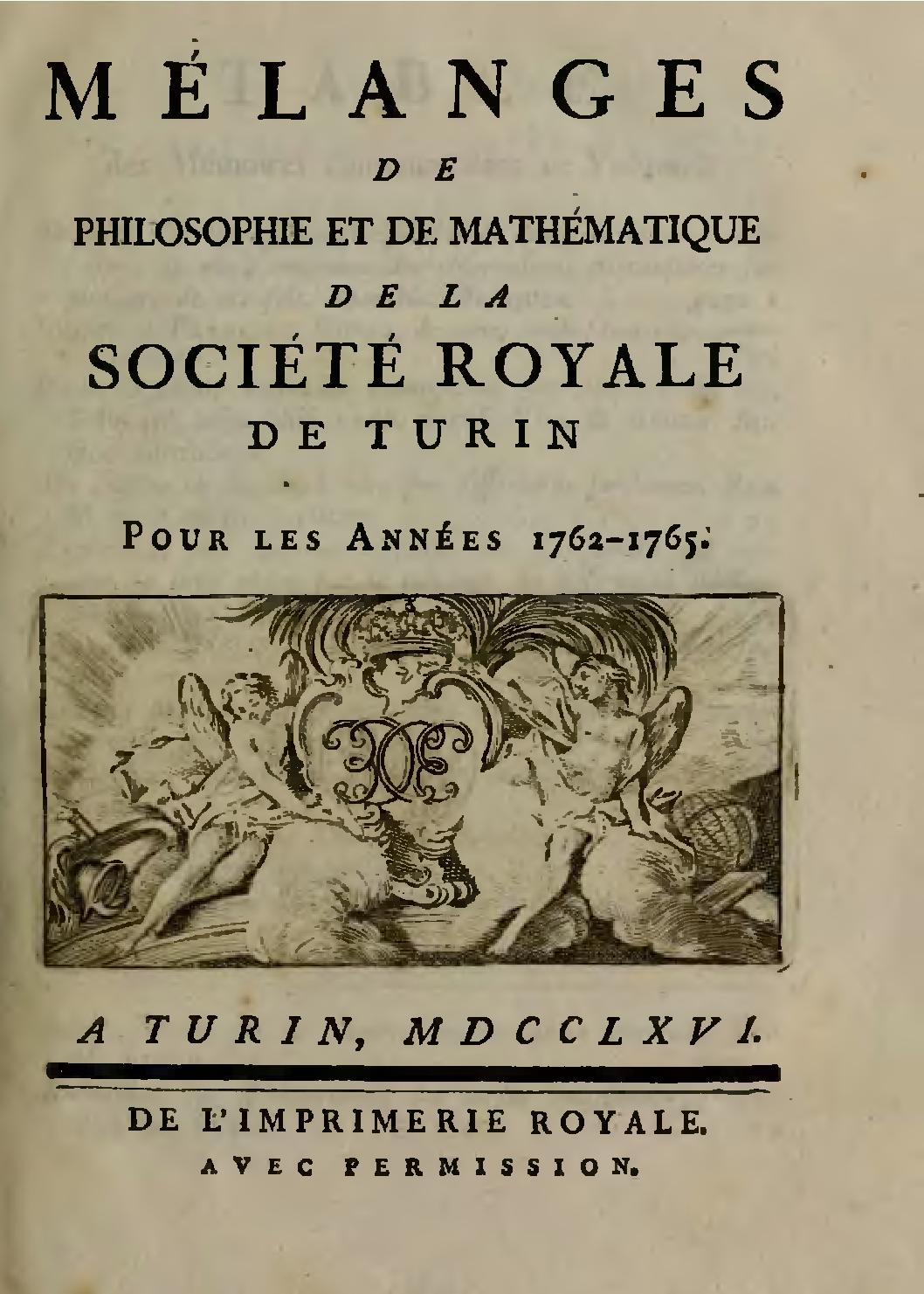}}

\href{http://archive.org/details/mlangesdephilo03soci}{\includegraphics[width=\marginparwidth]{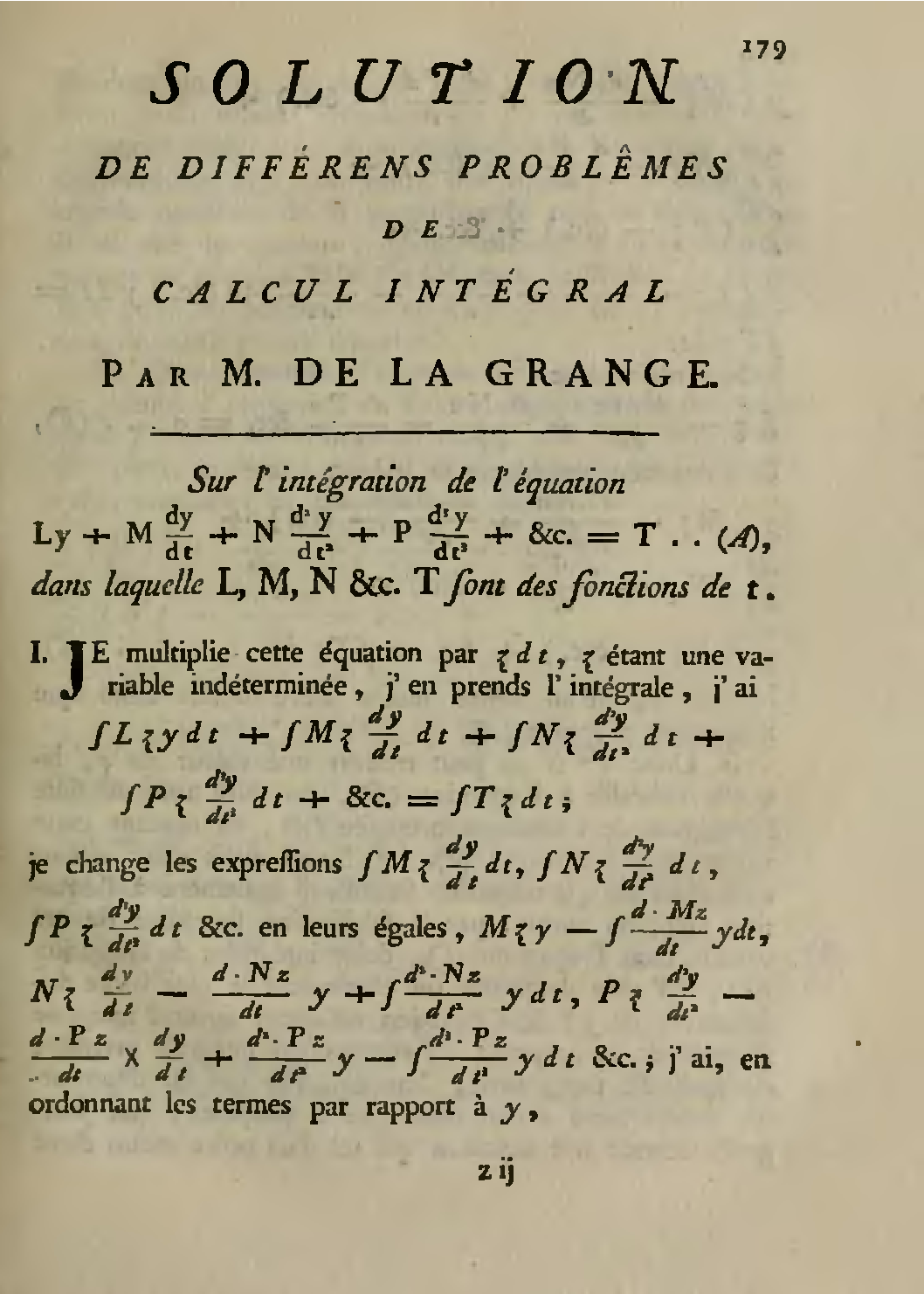}}
}
The first documented use of adjoint equations goes back to \cite{Lagrange}, who described a method to lower the order of a general linear ordinary differential equation and applied it to problems as diverse as fluid motion, vibrating strings and the orbit of planets.  His method used one or more particular solutions of a certain auxiliary equation obtained through integration by parts, which became known as ``\'equation adjointe'' in the French literature. The French term ``adjointe'' was a century later transliterated into the German ``adjungierte'' by \cite{Fuchs} and the English ``adjoint'' (after a brief alternation with ``adjunct'') by \cite{Forsyth} and \cite{Craig}.
``Self-adjoint'' problems (those for which the direct and adjoint equation coincide) became prominent with the development of the theory of integral equations and Hilbert spaces, and laid out the foundation for quantum mechanics.
\marginpar{
\vspace{1.4em}
\mptit{Courant--Hilbert} original title page

\centering
\href{http://gdz.sub.uni-goettingen.de/dms/load/toc/?PPN=PPN38067226X&IDDOC=264822}{\includegraphics[width=\marginparwidth,clip=true,trim=2.4cm 0cm 2.15cm 0.5cm]{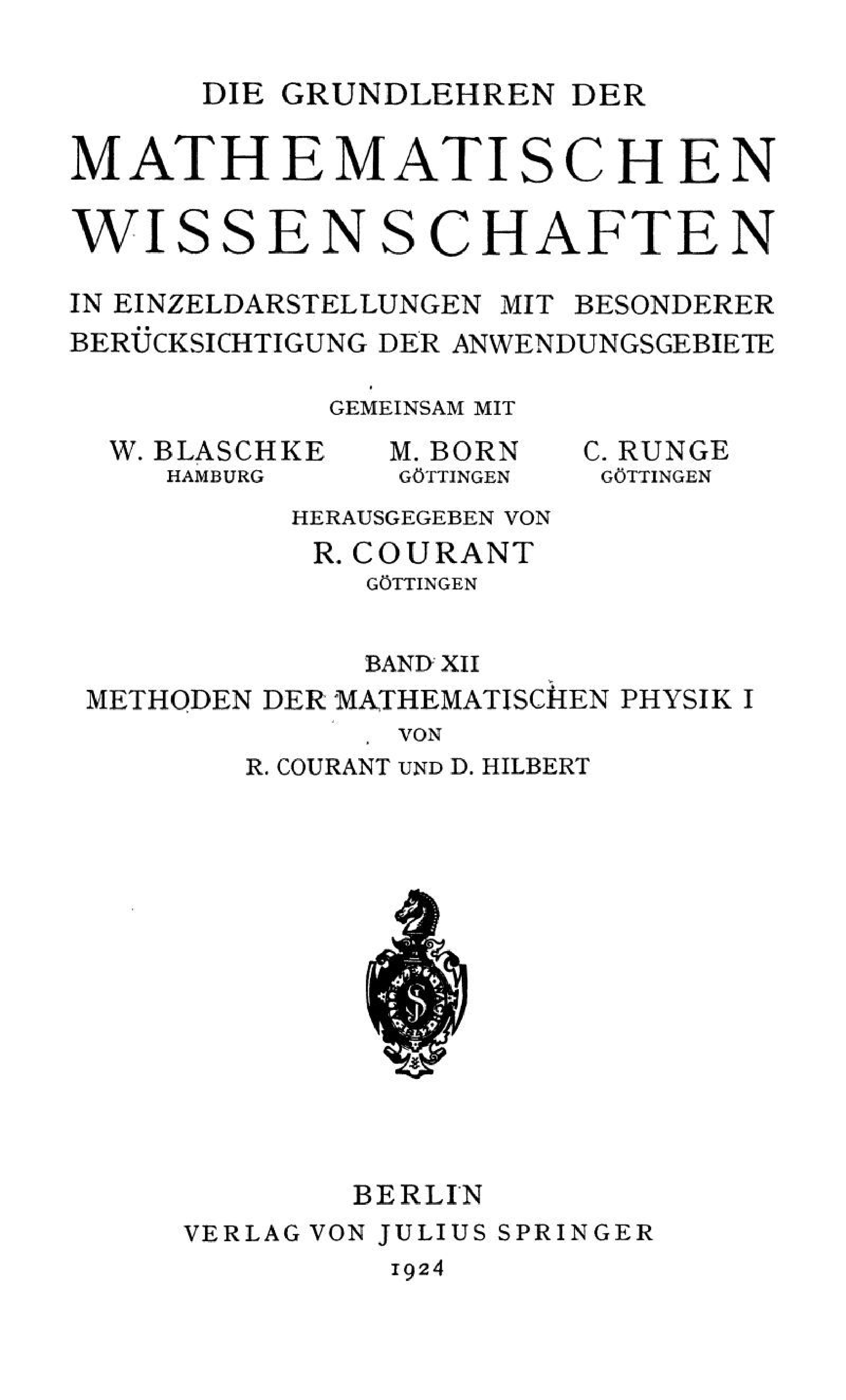}}}
The qualification ``adjungierte'' as used in \cite{Courant} mostly refers to Fredholm integral equations with mutually transposed kernels --- $K(t,s)$ being the transpose of $K(s,t)$ --- or to their approximation by mutually transposed matrices, the link with Lagrange's ``\'equation adjointe'' being that every reasonably well-behaved differential problem (equation and boundary conditions) can be reformulated as an integral equation and the corresponding adjoints coincide.

In the last half-century, with the advent of computers on one hand and control theory on the other, the adjoints of discrete problems, governed by transposed matrices, acquired an emphasis of their own. It became evident that, in addition to being a powerful analytical tool, the adjoint formulation held numerical value in reducing computation times for a variety of problems in mathematical physics. It also became evident that this value was available whether the original problem was self-adjoint or not. A large number of applications opened up
. In preparation we will now introduce adjoints of discrete and continuous systems of increasing complexity, thus reversing, as not rarely happens, the historical order.

\subsection{The simplest example: transpose of a matrix}\label{simplest}
Let us consider an \emph{instantaneous}, \emph{discrete} linear system (one in which a finite number of variables depend linearly on a finite number of others with no time delay) described by the explicit expression\footnote{The juxtaposition of two vectors (identified by bold small letters) or matrices (identified by bold capital letters) is used here to denote index contraction, with order significant:\; $\nvec{v}\nodot\nvec{u}=\sum_i v_i u_i$;\; $(\mat{A}\nodot\nvec{u})_i=\sum_j A_{ij}u_j$;\; $(\nvec{v}\nodot\mat{A})_j=\sum_i v_i A_{ij}$;\; $(\mat{A}\nodot\mat{B})_{ij}=\sum_k A_{ik} B_{kj}$.}
\be\nvec{x}=\mat{H}\nodot\nvec{u},
\ee{sdsystem}
$\nvec{x}$ being an array of $M$ real or complex variables (an $M$-vector), $\nvec{u}$ an $N$-vector, and $\mat{H}$ an $M\times N$-matrix.

If the quantity of interest (the objective) is not the entire state $\nvec{x}$ but a scalar linear function $J$ of it, expressed through the linear form
\be J=\nvec{y}\nodot\nvec{x}
\ee{sdobj}
($\nvec{y}$ being an $M$-vector of coefficients), the computation of $J$ would normally entail the sequential evaluation of \eqref{sdsystem}, requiring O$(MN)$ \emph{flops} (floating-point operations), and \eqref{sdobj}, requiring O$(M)$ flops.
However, substituting \eqref{sdsystem} into \eqref{sdobj} gives
\( \nvec{y}\nodot\nvec{x}=\nvec{y}\nodot\mat{H}\nodot\nvec{u}=\nvec{v}\nodot\nvec{u}
\)
where \(\nvec{v}=\nvec{y}\nodot\mat{H},
\)
or equivalently
\be\nvec{v}=\mat{H}^\trasp\nodot\nvec{y}, 
\ee{sdadjoint}
superscript $\,^\trasp$ denoting transposition.
Therefore the objective $J$ may also be rewritten in terms of the new $N$-vector $\nvec{v}$ as
\be J=\nvec{v}\nodot\nvec{u} .
\ee{sdobjsolved}

\begin{table}[b!]
\newlength\sidebarwidth
\newlength\sidebaroffs
\setlength\sidebaroffs\marginparsep
\addtolength\sidebaroffs\marginparwidth
\setlength\sidebarwidth\sidebaroffs
\addtolength\sidebarwidth\textwidth

\begin{minipage}{\sidebarwidth}
\small
\hrulefill
\subsubsection*{The hermitian adjoint}
A frequent alternate definition of the adjoint equation uses an inner product rather than a linear form. While we do not use an inner product in this article, elucidating the relationship may ease comprehension.

Functional analysis (\eg\ \cite{Dunford}) distinguishes between a \emph{Banach space}, a vector space of functions endowed with a topology that allows the concepts of continuity and completeness to be defined, and a \emph{Hilbert space}, a Banach space with the addition of an inner product
. Using the algebraic properties of a vector space alone, the notion of its \emph{dual} is introduced as the space of all linear functions from a vector to a scalar. The dual and original vector space do not generally overlap: for instance, the dual of a space of continuous and integrable functions may be a space of distributions (the mathematically orthodox way to deal with the Dirac delta function). Adjoint problems properly belong in the dual space. On the other hand when an inner product exists, which we may denote as $s=\left<\nvec{x}_1,\nvec{x}_2\right>$, it establishes a mapping between the vector $\nvec{x}_1$ and a linear function from the vector $\nvec{x}_2$ to the scalar $s$, \ie\ between the original state space and its dual. It then becomes possible to argue whether the problem possibly coincides with its adjoint (is \emph{self-adjoint}) under this mapping.

For finite-dimensional vectors, such as appear in all numerical approximations, an element of the dual space is a linear form like \eqref{sdobj}; a general inner product can be defined as  \[
\left<\nvec{x}_1,\nvec{x}_2\right>=\nvec{x}_1^*\mat{Q}\nvec{x}_2
\]
where $\,^*$ denotes complex conjugation and $\mat{Q}$ is an arbitrary positive-definite (hermitian) matrix. \eqref{sdobj} and \eqref{sdobjsolved} may then be alternately written as
\(
J=\left<\overbar{\nvec{y}},\nvec{x}\right>=\left<\overbar{\nvec{v}},\nvec{u}\right>,
\) 
where
\[
\nvec{y}=\overbar{\nvec{y}}^*\mat{Q},\qquad \nvec{v}=\overbar{\nvec{v}}^*\mat{Q}.
\]

Both $\overbar{\nvec{v}}$ and $\nvec{v}$ are suitable definitions of an adjoint vector, which can be distinguished by naming $\overbar{\nvec{v}}$ the \emph{hermitian adjoint}, and are easily converted to one another where both apply.
The hermitian adjoint definition is convenient when the emphasis is on self-adjoint problems (\S\ref{selfadjoint}), since a mapping between the state space and its dual (and the choice of a specific $\mat{Q}$) is needed for such a notion to make sense. On the other hand the definition, used everywhere in this article, of adjoint equation as the one yielding the sensitivity $\nvec{v}$ applies to a larger class of linear problems (including those that are not posed in a Hilbert space) and naturally extends to nonlinear ones (\S\ref{nonlinear}). In computation it dispenses with matrix $\mat{Q}$ and a few complex conjugations.\\

\hrulefill
\end{minipage}
\end{table}

The computational interest of this seemingly trivial relationship is that one can calculate $\nvec{v}$ from $\nvec{y}$ just once, at the 
same O$(MN)$ cost as a single application of \eqref{sdsystem}, and then obtain $J$ for a number of different inputs at a cost of O$(N)$ flops for each. The coefficient $\nvec{v}$, which in this linear example provides the derivative of the objective $J$ with respect to the input $\nvec{u}$, may be named \emph{sensitivity} of the objective to the input.
Equation \eqref{sdadjoint}, which we shall name the \emph{adjoint problem}
of \eqref{sdsystem}, therefore provides this sensitivity, at a computational advantage with respect to applying \eqref{sdsystem} to each input in turn.

More generally, one may be interested in computing $P$ objective scalars $J_p$ for $Q$ different input vectors $\nvec{u}_q$ (with both $P\le M$ and $Q\le N$ for them to be linearly independent).
\marginpar{\mptit{The adjoint advantage}: with fewer inputs than outputs direct computation is faster, with fewer outputs than inputs adjoint computation is.}
The important balance to be kept in mind is that the advantage of using an adjoint depends on whether there are more inputs or outputs. In this example it's easy to work out that if $P/M<Q/N$, computing the adjoint first is faster, whereas if $P/M>Q/N$ the direct system gives a quicker answer.

\subsection{What are adjoints used for?}
\label{What}
As the previous example has shown, adjoint equations yield sensitivity fields. In the following sections we shall consider several variations of discrete and continuous systems and define an adjoint problem for each. The common thread will be that for every linear problem one can, at least in principle, define an input-output relationship like \eqref{sdsystem} mediated by a matrix (or, in a continuous setting, by an operator) $\mat{H}$, named its \emph{resolvent}, and all the adjoint problems we shall define are, in this sense, equivalent to \eqref{sdadjoint}. However, explicitly calculating the resolvent may not always be a practical thing to do. Just as replacing \eqref{sdsystem}--\eqref{sdobj} by \eqref{sdadjoint}--\eqref{sdobjsolved} involves a computational economy when fewer outputs than inputs are used, calculating the adjoint of each individual step that defines the system is generally more efficient than calculating an all-encompassing resolvent and applying \eqref{sdsystem}--\eqref{sdadjoint} to it. This will become clearer as we add more examples along the way. 

The adjoint formulation is useful when one is seeking to obtain one or a few outputs of a system for a wide range of possible inputs. There are several such cases in fluid mechanics (and other disciplines), in particular in stability theory as will be expounded in this article, but the greatest advantage is obtained in optimization. In fact, the typical optimization problem has a single objective function (possibly combining multiple objectives through suitable weights) that has to be minimized or maximized with respect to a large number, or even a continuum, of input variables: a perfect application for adjoints!

\subsection{Adjoint of a system of linear algebraic equations}
Let us now examine an instantaneous linear system where the relationship between state $\nvec{x}$ and input $\nvec{u}$ is implicit, namely of the form
\be\mat{A}\nodot\nvec{x}=\nvec{u}
\ee{isdsystem}
with $\mat{A}$ denoting a non-singular $M\times M$ square matrix.
One obvious way to devise an adjoint of this problem is to convert \eqref{isdsystem} to \eqref{sdsystem} by using the inverse $\mat{H}=\mat{A}^{-1}$ and then proceed as in \S\ref{simplest}; however a more powerful alternative is to write the objective $J$ defined by \eqref{sdobj} in the form \eqref{sdobjsolved}, with an as yet unknown coefficient vector $\nvec{v}$, and then substitute \eqref{isdsystem} into it:
\be J =\nvec{v}\nvec{u}=\nvec{v}\mat{A}\nvec{x}= \nvec{y}\nvec{x}.
\ee{vmult}
It is evident by inspection that this becomes an identity for any $\nvec{x}$ (and consequently for any $\nvec{u}$) if
\be\mat{A}^{\!\trasp}\nvec{v}=\nvec{y},
\ee{isdadj}
which deserves to be named the adjoint system of \eqref{isdsystem}. The advantage is once again computational: obtaining $\nvec{v}$ from \eqref{isdadj} only requires the solution of a linear system with a specific \rhs, which is classically known to be a less costly numerical algorithm than constructing the inverse $\mat{H}=\mat{A}^{-1}$ (and even more so if $\mat{A}$ happens to be a sparse or banded matrix).

The adjoint of a system of linear equations also finds a useful role in defining the \emph{compatibility condition}
that must be satisfied when the coefficient matrix is singular, for the system to admit a solution: if $\mat{A}$'s determinant is zero, both the direct and 
\marginpar{\mptit{Compatibility condition}: for an inhomogeneous singular linear system to admit a solution, the product of its r.h.s. and its adjoint homogeneous solution must be zero.}
transposed systems have a nonzero homogeneous solution, as is known from linear algebra, and system \eqref{isdsystem} only admits solutions under a suitable compatibility condition. If $\nvec{v}$ is any nonzero solution of the homogeneous adjoint system, $\nvec{v}\mat{A}=0$, contracting both sides of \eqref{isdsystem} with $\nvec{v}$ gives
\be\nvec{v}\nvec{u}=0
\ee{compatib}
which is thus a nontrivial necessary condition on $\nvec{u}$ for the solution $\nvec{x}$ to exist.

\subsection{Entering dynamics: the adjoint of a discrete-time dynamical system}\label{discrete-time}
Let us now complicate matters slightly by adding dynamics (\ie, time dependence) in its simplest form, that of a discrete-time dynamical system described by the state-transition equation
\marginpar{\mptit{Discrete-time dynamical system}: system characterized by a numerable sequence of states each of which is completely determined by its predecessor.}
\be \nvec{x}_{n+1}=\mat{A}_n \nodot\nvec{x}_n
\ee{dtsystem}
with state dimension $M$, $M\times M$ transition matrix $\mat{A}_n$, initial condition $\nvec{x}_0$, discrete time $n$ running from $0$ to $N-1$, and an objective depending linearly on $\nvec{x}_N$:
\be J=\nvec{y}\nvec{x}_N.
\ee{dtobj}
The direct computation of $J$ involves O$(NM^2)$ flops for \eqref{dtsystem} and O$(M)$ flops for \eqref{dtobj}. We can also unroll the process in the form of the explicit product chain
\[J=\nvec{y}(\mat{A}_{N-1}\nodot(\mat{A}_{N-2}\nodot( \cdots \mat{A}_{1}\nodot(\mat{A}_{0}\nodot\nvec{x}_0)))).
\]
Now, let's associate the products to the left rather than to the right:
\[J=((((\nvec{y}\mat{A}_{N-1})\nodot\mat{A}_{N-2} \cdots)\nodot \mat{A}_{1})\nodot\mat{A}_{0})\nodot\nvec{x}_0.
\]
Clearly we have again obtained a compact expression of the objective as a function of the input, namely the linear form
 $J=\nvec{v}_0\nvec{x}_0$ where
\be
\nvec{v}_{\!n}=\mat{A}_n^{\!\trasp}\nvec{v}_{\!n+1}
\ee{dtadjoint}
and $\nvec{v}_{\!N}=\nvec{y}$. Equation \eqref{dtadjoint} defines for all purposes a new dynamical system, which can be denoted as adjoint of \eqref{dtsystem}, with the peculiarity that discrete time $n$ runs backwards from $N-1$ down to $0$, and that instead of an initial it has a ``terminal'' condition $\nvec{v}_{\!N}=\nvec{y}$. Notice also that with no difficulty the transition matrix $\mat{A}$ has been allowed to change with $n$; time-invariant systems are of course included as a special case.

Just as in \S\ref{simplest}, the computation of the objective $J$ for several different initial conditions $\nvec{x}_0$ may be obtained by applying a single sequence of \eqref{dtadjoint}, at the same cost as a single sequence of \eqref{dtsystem}, and then calculating $J=\nvec{v}_0\nvec{x}_0$ at a cost of O$(M)$ flops for each initial condition. This procedure only involves matrix-vector products, and is much less computationally expensive than calculating the resolvent $\mat{H}=\mat{A}_{N-1}\nodot\mat{A}_{N-2} \cdots\nodot \mat{A}_{1}\nodot\mat{A}_{0}$ through matrix-matrix products.
In addition,
the same value of $J$ repeatedly appears at every intermediate step of \eqref{dtsystem} and \eqref{dtadjoint}:
\be
J=\nvec{v}_{\!n}\nvec{x}_n = \text{const.}\quad \text{for } n=0\ldots N ;
\ee{fundid}
this property turns up very useful when testing an adjoint computer program (\S\ref{AdjointProgram}).

The evolution equation
of a more general dynamical system with a time-varying input term $\nvec{u}_n$ weighted by an input matrix $\mat{B}_n$ may be written as
\marginpar{\mptit{General dynamical sys\-tem} with a time-varying input term and an objective depending on the state at all times.}
\be \nvec{x}_{n+1}=\mat{A}_n \nodot\nvec{x}_n + \mat{B}_n\nodot \nvec{u}_n\qquad(n=0\ldots N-1).
\ee{gdtsystem}
Correspondingly, a more general objective depending on the state at all times is
\be J\, = \sum_{n=0}^{N} \nvec{y}_{\!n} \nvec{x}_n .
\ee{gdtobj}

Obtaining an adjoint for this more general case is no more difficult than in the previous example. It only requires recursively eliminating (in the sense of Gaussian elimination) each $\nvec{x}_{n+1}$ for $n=N-1\dots 0$. First, inserting $\nvec{x}_N$ from \eqref{gdtsystem} into \eqref{gdtobj} gives
\begin{multline*} J = \nvec{y}_{\!N}(\mat{A}_{N-1}\nodot \nvec{x}_{N-1} + \mat{B}_{N-1}\nodot \nvec{u}_{N-1}) +  \sum_{n=0}^{N-1} \nvec{y}_{\!n} \nvec{x}_n =\\=
 \nvec{v}_{\!N} \mat{B}_{N-1} \nvec{u}_{N-1} + \nvec{v}_{\!N-1} \nvec{x}_{N-1} +   \sum_{n=0}^{N-2} \nvec{y}_{\!n} \nvec{x}_n
\end{multline*}
where $\nvec{v}_{\!N}=\nvec{y}_{\!N}$ and
$ \nvec{v}_{\!N-1} = \mat{A}_{N-1}^{\!\trasp} \nvec{v}_{\!N}+\nvec{y}_{\!N-1}.
$
By recursion
\be J\, = \sum_{n=0}^{N-1} (\nvec{v}_{\!n+1} \mat{B}_n)\nodot \nvec{u}_n + \nvec{v}_0 \nvec{x}_0
\ee{gdtobjsolved}
where
\be \nvec{v}_{\!n} = \mat{A}_{n}^{\!\trasp} \nvec{v}_{\!n+1}+\nvec{y}_{\!n} \qquad(n=N-1\ldots0) .
\ee{gdtadjoint}
This is the equation adjoint to \eqref{gdtsystem}.

An alternate way to derive the same result is to observe that \eqref{gdtsystem} and \eqref{gdtobj} may be formally reduced to a particular case of \eqref{dtsystem} and \eqref{dtobj} by defining an \emph{augmented state} that includes in a single much larger vector the components of all states and all inputs at all times. If sparseness of the transition matrices is suitably taken into account, \eqref{gdtobjsolved} and \eqref{gdtadjoint} are obtained again.

\subsection{From discrete to continuous time}
\label{contadj}
The adjoint of a system described through differential equations can also be described through differential equations, and the constructive procedure discovered in his time by Lagrange allows us to derive such adjoint equations.

Let us consider the linear, continuous-time dynamical system
\marginpar{\mptit{Continuous-time} dynamical system and objective.}
\be \frac{\d\nvec{x}}{\d t}=\mat{A} \nodot\nvec{x} + \mat{B}\nodot \nvec{u}
\ee{ctsystem}
with initial condition $\nvec{x}(0)=\nvec{x}_0$ and objective
\be J\;=\int_0^T \nvec{y}(t)\nvec{x}(t) \d t.
\ee{ctobj}
Useful information about the sensitivity 
of $J$ to the initial condition 
and input 
may be gathered by looking at a discrete approximation of \eqref{ctsystem}.

The most basic (first-order, explicit) numerical discretization of \eqref{ctsystem} with a stepsize $\Delta t=T/N$ is a discrete dynamical system of the general form \eqref{gdtsystem}, namely
\be \nvec{x}_{n+1}=\left(1+\Delta t\,\mat{A}\right) \nodot\nvec{x}_n + \Delta t\,\mat{B} \nvec{u}_n. 
\ee{dctsystem}
The corresponding adjoint equation, according to \S\ref{discrete-time}, is
\be \nvec{v}_{\!n} = (1+\Delta t\,\mat{A}^{\!\trasp})\, \nvec{v}_{\!n+1}+\Delta t\,\nvec{y}_{\!n}, 
\ee{dctadjoint}
to be marched backwards in $n$ from the terminal condition $\nvec{v}_{\!N}=0$. The look of equation \eqref{dctadjoint} suggests that it might be the numerical discretization of the new differential equation
\be-\frac{\d\nvec{v}}{\d t}=\mat{A}^{\!\trasp} \nvec{v}+\nvec{y}.
\ee{ctadjoint}

That this is indeed the adjoint of \eqref{ctsystem} can be proved in a similar manner as \eqref{vmult} was derived. We first build a linear form by contracting both sides of \eqref{ctsystem} with an as yet unknown vector-valued function $\nvec{v}(t)$ and integrating:
\[ \int_0^T \nvec{v} \frac{\d\nvec{x}}{\d t} \d t\;=\int_0^T \nvec{v} \left(\mat{A} \nodot\nvec{x} + \mat{B}\nodot \nvec{u}\right) \d t.
\]
This identity may then be led to only contain $\nvec{x}$ and not its derivative through an integration by parts:
\be
\Big[\nvec{v} \nvec{x}\Big]_0^T - \int_0^T\frac{\d\nvec{v}}{\d t}\nodot\,\nvec{x} \d t\; = \int_0^T \nvec{v} \left(\mat{A} \nodot\nvec{x} + \mat{B}\nodot \nvec{u}\right) \d t,
\ee{byparts}
whence, by collecting the coefficients of $\nvec{x}$ and comparing with \eqref{ctobj},
\be J\;=\int_0^T\left(-\,\frac{\d\nvec{v}}{\d t}-\nvec{v}\mat{A} \right)\nvec{x}(t)\d t=
\nvec{v}(0)\, \nvec{x}(0)+\int_0^T \nvec{v} \mat{B}\nodot \nvec{u} \d t
\ee{ctsolvedobj}
provided $\nvec{v}$ is chosen to be a solution of \eqref{ctadjoint} with the terminal condition $\nvec{v}(T)=0$.
That differential equation for $\nvec{v}$, to be integrated backwards in time just like \eqref{dctadjoint}, thus provides the explicit expression of the objective as a linear function of the initial condition and input data and therefore deserves to be called the adjoint equation of \eqref{ctsystem}.

Remarkably the adjoint equation is endowed with a similar structure and computational difficulty as the original direct problem, as clearly appears from the comparison of the corresponding discretizations \eqref{dctsystem} and \eqref{dctadjoint}. Just like the discrete systems of the previous section, \eqref{ctadjoint} must be solved by marching backwards in time with the terminal condition%
\footnote{A nonzero terminal condition may optionally be introduced to account for the presence of an additional term proportional to $\nvec{x}(T)$ in \eqref{ctobj}. Alternately, $\nvec{v}(T)$ may be assumed to be always zero and such a term included in $\nvec{y}$ as a delta function.}
$\nvec{v}=0$ at $t=T$, in order to render the expression \eqref{ctsolvedobj} of $J$ independent of $\nvec{x}(T)$ and a function of $\nvec{x}(0)$ and $\nvec{u}$ only. In addition, the relationship between the direct and adjoint problem's eigen\-values, which will be seen in \S\ref{bvpevp}, is such that if the direct problem is well-posed in forward time (for instance, it is the semi-discretization of a parabolic problem with a forward direction of stable evolution), the adjoint problem is well-posed in backward time.

\stabilityref{
In the context of stability and receptivity analysis, early applications of the adjoint equation \eqref{ctadjoint} to a parabolic problem can be found in \cite{Luchini1bis,Luchini1}, who calculated the receptivity of the G\"ortler instability, and \cite{Airiau2}, who developed an adjoint parabolized stability equation.  
}

\subsection{Connection between adjoint and resolvent}
The solution of the linear system \eqref{ctsystem}, with initial condition $\nvec{x}(0)=\nvec{x}_0$ and input $\nvec{u}(t)$, may always be formally written as a linear combination of homogeneous solutions, \ie\, in matrix notation,
\be\nvec{x}(t)=\mat{H}(t,0)\nvec{x}_0+\int_0^t\mat{H}(t,\tau)\mat{B}(\tau)\nodot \nvec{u}(\tau) \d\tau
\ee{propag}
where $\mat{H}(t,\tau)$ is the solution in the $[\tau,t]$ interval of the homogeneous matrix differential equation
\marginpar{\mptit{Resolvent}, also known as Green's function or propagator.}
\be\frac{\d\mat{H}}{\d t}=\mat{A}(t) \nodot\mat{H}(t,\tau)
\ee{resolvdir}
with initial condition\footnote{$\;\delta_{ij}$ being Kronecker's symbol, \ie\ $1$ for $i=j$ and $0$ for $i\neq j$.} $H_{ij}=\delta_{ij}$ at $t=\tau$. $\mat{H}(t,\tau)$ is by definition the resolvent introduced in \S\ref{What} (also known as \emph{Green's function} or \emph{propagator}) of \eqref{ctsystem}.

Inserting \eqref{propag} into the definition of the objective \eqref{ctobj} (while for simplicity assuming $\nvec{x}_0=0$ and $\mat{H}(t,\tau)=0$ for $t<\tau$) gives
\[ J\;=\int_0^T \int_0^T\nvec{y}(t)\mat{H}(t,\tau)\mat{B}(\tau)\nodot \nvec{u}(\tau) \d\tau \d t;
\]
reversing the order of integration and comparing to \eqref{ctsolvedobj} then yields
\be\nvec{v}(\tau)\;= \int_\tau^T \nvec{y}(t)\mat{H}(t,\tau) \d t,
\ee{adjfromres}
account having been taken that $\mat{H}(t,\tau)=0$ for $t<\tau$. It may be noticed that \eqref{propag} is analogous to \eqref{sdsystem}, and \eqref{adjfromres} to \eqref{sdadjoint}, of \S\ref{simplest}.

It follows that the transposed resolvent $\mat{H}^\trasp(t,\tau)$, if regarded as a function of its second argument $\tau$ in the $[0,t]$ interval, is the resolvent of the adjoint problem and obeys the homogeneous adjoint differential equation
\be-\,\frac{\d\mat{H}^\trasp}{\d \tau}=\mat{A}^{\!\trasp}(\tau)\mat{H}^\trasp(t,\tau),
\ee{resolvadj}
to be integrated backwards from $\tau=t$ to $\tau=0$ with the terminal condition $H^\trasp_{ij}=\delta_{ij}$ at $\tau=t$.

As an easy to remember conclusion, the resolvent obeys in symmetrical fashion the homogeneous direct equation as a function of final time $t$ and the homogeneous adjoint equation as a function of initial time $\tau$. However, as the previous examples have shown, the resolvent matrix is mostly a useful theoretical tool whereas calculating either the direct or the adjoint vector is computationally faster.

\subsubsection{The resolvent of a time-invariant problem}
When $\mat{A}$ is independent of time, the resolvent is classically represented as\footnote{\ $\otimes$ denoting the dyadic product of two vectors to produce a matrix.}
\be\mat{H}(t,\tau)=\exp\left[\mat{A}\;(t-\tau)\right]\equiv\sum_{m=1}^{M}e^{\sigma_m\, (t-\tau)}\,\nvec{u}_m\otimes\nvec{v}_m
\ee{eigenexpansion}
where $\sigma_m$ are $\mat{A}$'s eigen\-values (assumed to be distinct for simplicity here) and $\nvec{u}_m$, $\nvec{v}_m$ are $\mat{A}$'s right and left eigen\-vectors. That \eqref{eigenexpansion} is a solution of both \eqref{resolvdir} and \eqref{resolvadj} should be self-evident, as $\sigma_m$ are the eigen\-values of $\mat{A}$ and $-\sigma_m$ the eigen\-values of $-\mat{A}^{\!\trasp}$.

\stabilityref{
The sum in \eqref{eigenexpansion} can be cut down to a single dominant mode when the latter is exponentially amplified. As such, it plays an important role in the temporal receptivity analysis of global modes \citep{Hill1} and in the spatial receptivity analysis of parallel flows \citep{Fedorov,Tumin3,Hill2}. Its extension through an inhomogeneous multiple-scale WKBJ approximation provides the receptivity of quasi-parallel flows \citep{ZuccherT,Zuccher3}. It must be remarked that this single-mode approximation (very common in the investigation of boundary-layer instabilities) is the only example in this review where computing the resolvent is computationally faster than resorting to adjoint equations.
}

\subsection{Adjoint of a boundary-value problem or of an eigen\-value problem}\label{bvpevp}
\marginpar{\mptit{Adjoint of a boundary- value problem}: specify $M-P$ conditions at one end and $P$ conditions at the other end in such a way that the explicit expression of the objective only contains the original $M$ boundary data.}
It may have been noticed that the finite term arising from the integration by parts in \eqref{byparts} contains exactly twice as many addends as there are initial conditions for $\nvec{x}$, and that the terms not used by such initial conditions have been eliminated by choosing the terminal conditions for $\nvec{v}$ appropriately. In fact this is a rule:
for a more general boundary-value problem ({b.v.p.}) specifying $P$ conditions at $t=0$ and $M-P$ conditions at $t=T$ (where the independent variable $t$ does not necessarily have to represent time) it is always possible to choose $M-P$ conditions at $t=0$ and $P$ conditions at $t=T$ for $\nvec{v}$ in such a way that \eqref{ctsolvedobj} only contains the original boundary data. Therefore the adjoint of a {b.v.p.} will be another {b.v.p.}

An additional complication is that the solution of a {b.v.p.}, contrary to the initial-value problem to which Picard's classical existence and uniqueness theorem applies, cannot \emph{a priori} be granted to exist and/or be unique. Indeed another fruitful application of adjoints, although one that will not be addressed here, is precisely in devising existence and uniqueness proofs for such solutions.

If the {b.v.p.} is homogeneous and contains a parameter $\sigma$, it becomes an eigen\-value problem, one that generally admits only the null (trivial) solution except for special values of $\sigma$ which are named its \emph{eigen\-values}. Its adjoint problem (equation and boundary conditions) can be determined just as for any other {b.v.p.}, and will depend on $\sigma$ too. Under reasonable regularity conditions (essentially requiring that a so called \emph{normal form} --- a form explicit in the state derivatives such as \eqref{ctsystem} --- exists and is not singular, \eg\ \cite{Tricomi}), differential eigen\-value problems share the same properties as matrix eigen\-value problems. In particular, the direct and adjoint problem have the same eigen\-values and mutually orthogonal eigen\-functions.

For a (generalized) matrix eigen\-value problem of the form
\be\mat{A}\nvec{x}=\sigma\mat{B}\nvec{x},
\ee{mateigv}
with adjoint
\be\nvec{v}\mat{A}=\sigma\nvec{v}\mat{B},
\ee{matadjeigv}
the existence of a nontrivial solution of \eqref{matadjeigv} for the same $\sigma$ where one of \eqref{mateigv} exists is a consequence of the fact that the direct and transposed matrix have the same determinant.
Orthogonality is classically proven by considering an eigen\-vector $\nvec{x}_1$ of \eqref{mateigv} corresponding to an eigen\-value $\sigma_1$ and an adjoint eigen\-vector $\nvec{v}_2$ of \eqref{matadjeigv} corresponding to a different eigen\-value $\sigma_2$. Contracting \eqref{mateigv} with $\nvec{v}_2$ and \eqref{matadjeigv} with $\nvec{x}_1$, and subtracting, gives
\[\nvec{v}_2(\mat{A}\nvec{x}_1)-(\nvec{v}_2\mat{A})\nvec{x}_1=\nvec{v}_2(\sigma_1\mat{B}\nvec{x}_1)-(\sigma_2\nvec{v}_2\mat{B})\nvec{x}_1
\]
\ie
\[0=(\sigma_1-\sigma_2)\nvec{v}_2\mat{B}\nvec{x}_1
\]
whence $\nvec{v}_2\mat{B}\nvec{x}_1=0$ (by a similar token, also $\nvec{v}_2\mat{A}\nvec{x}_1=0$) follows unless $\sigma_1=\sigma_2$. Furthermore, if the 
eigen\-vectors form a complete set (a \emph{basis}) and $\nvec{v}_i\mat{B}\nvec{x}_i \neq 0$ for all $i=1\ldots M$, an arbitrary 
vector $\nvec{x}$ can be expanded as  
\be
\nvec{x}\,=\sum_{i=1}^{M} \frac{\nvec{v}_i\mat{B}\nvec{x}}{\nvec{v}_i\mat{B}\nvec{x}_i}\nvec{x}_i .
\ee{eigvexp}
\marginpar{\mptit{Projector}: the adjoint eigen\-vector multiplied by $\mat{B}$ projects an arbitrary vector onto the corresponding direct eigen\-vec\-tor.}
This formula may be read to say that, with proper normalization, $\nvec{v}_i\mat{B}$ is the \emph{projector} that extracts the component of $\nvec{x}$ along $\nvec{x}_i$.

For a differential eigen\-value problem in normal form, similar properties are proven by first reformulating it as a Fredholm integral equation governed by a compact operator (this is where the normal form is required) and then showing that a compact operator can be uniformly approximated by a matrix to any desired precision. 
In addition, if the problem is self-adjoint (see below) the set of eigen\-vectors is complete 
\citep{Courant,Tricomi}. Notice, however, that if boundary conditions are given over an infinite interval (the operator is no longer compact) completeness may be lost or may involve continuous ranges of eigen\-values (a \emph{continuous spectrum}).

\subsubsection{Self-adjoint problems}
\label{selfadjoint}
When equations \eqref{mateigv} and \eqref{matadjeigv} coincide (or, for complex-valued matrix elements, when the second is the complex conjugate of the first, denoted by a $^*$ superscript), the problem is qualified \emph{self-adjoint}. This occurs when both matrices $\mat{A}$ and $\mat{B}$ are hermitian: $\mat{A}^{\!\trasp}=\mat{A}^*$ and  $\mat{B}^\trasp=\mat{B}^*$. Indeed, in many differential-equations and mathematical-physics textbooks
adjoint problems are introduced mostly in preparation for the special properties of self-adjoint ones, namely:

\marginpar{\mptit{Self-adjoint problem}:

\noindent problem that coincides with its adjoint in a suitable inner-product space.}
\begin{enumerate}
\item for each eigen\-vector $\nvec{x}_n$,  $\nvec{v}_{\!n}=\nvec{x}_n^*$;
\item for each eigen\-value $\sigma_n$, $\sigma_n=\sigma_n^*$ (all eigen\-values are real);
\item with suitable normalization, eigen\-vectors are orthonormal with respect to the hermitian inner product $\nvec{x}_m^*\mat{B}\nvec{x}_n=\delta_{mn}$ (in the case of functional spaces, to the $\mathcal{L}_2$ inner product $\int_0^T\nvec{x}_m^*\mat{B}\nvec{x}_n \d t=\delta_{mn}$);
\item if $\mat{B}$ is non-singular, the set of eigen\-vectors always forms a basis.
\end{enumerate}
For instance, every second-order linear differential equation can be put in a \emph{Sturm-Liouville} form \citep{Tricomi}, which makes it self-adjoint provided the inner product is defined through a suitable weight function.

Here we shall mostly dispense with these properties, as this review purports to highlight the use of adjoints in the general non-self-adjoint case. 
A theoretical detail, however, has to be highlighted: whereas the definition of adjoint equation as the one providing a sensitivity only involves a linear form, the idea of a self-adjoint problem requires the additional existence of an inner product, 
so that direct and adjoint eigen\-functions live in the same Hilbert space and can be compared.
The inner product may be defined in more than one way, \eg\ with the intervention of a weight function as in Sturm-Liouville problems, and a problem that is self-adjoint under a certain inner product will not be so under another. Some problems may even turn out to be self-adjoint under an operation that lacks the algebraic properties of an inner product (an example will appear in \S\ref{ADA}).


\subsection{Structural sensitivity of an eigen\-value problem}
\label{structsens}
As a particular application of the adjoint, the objective whose sensitivity is sought may be chosen to be the eigen\-value itself. This will become useful every time the growth rate of an instability is being investigated as a function of the physical parameters of interest
.

In the context of the finite-dimensional system \eqref{mateigv} we can consider how the eigen\-value changes under an infinitesimal perturbation $\d \mat{A}$ of matrix $\mat{A}$. This is named a \emph{structural perturbation}, meaning a
\marginpar{\mptit{Structural perturbation}: perturbation of the coefficients rather than the forcing of a linear system.}
perturbation of the \emph{structure} (the coefficients) rather than the forcing (the \rhs ) of a linear system.
Taking the differential of both sides of \eqref{mateigv} gives 
\[
\d[ (\mat{A}-\sigma\mat{B} ) \,\nvec{x}]=(\d\mat{A}-\d\sigma\mat{B}) \,\nvec{x}+( \mat{A}-\sigma\mat{B}) \, \d \nvec{x}=0.
\]
The eigen\-vector perturbation $\d \nvec{x}$ can be algebraically eliminated from this equation by contracting it with the adjoint eigen\-vector $\nvec{v}$. Recalling \eqref{matadjeigv} in fact provides
\begin{equation}
\d \sigma=\dfrac{\nvec{v} \, \d \mat{A}\, \nvec{x}}{\nvec{v} \, \mat{B}\, \nvec{x}}.
\label{struct_sens}
\end{equation}
The reader may notice that this formula is also encountered in numerical matrix algebra \citep{Golub}, where it is used to define the condition number of an eigen\-value through its sensitivity to roundoff errors.

If the perturbation $\d \mat{A}$ has only one nonzero element, say $\d A_{mn}$, \eqref{struct_sens} can be further simplified to give      
\be
\d \sigma = \dfrac{v_{m}x_{n} }{\nvec{v} \, \mat{B}\, \nvec{x}}\,\d A_{mn}
.
\ee{eigensens}
This expression shows that the sensitivity of the eigen\-value to an infinitesimal disturbance $\d A_{mn}$ of an individual matrix element is proportional to the product between the components $x_{n}$ and $v_{m}$ of, respectively, the direct and adjoint eigen\-vector.

\stabilityref{
Equation \eqref{struct_sens} is the basis for the concepts of base-flow sensitivity and optimal base-flow perturbation of a stability problem \citep{Bottaro2}. Equation \eqref{eigensens} was used to introduce the structural-sensitivity map of a global mode as a means to determine its \emph{wavemaker} \citep{Giannetti03,Giannetti1}.
}

\subsection{Adjoints in optimization and control}
\label{optimization}
The simplifying power of adjoints is not limited to linear objectives. They reappear naturally in the optimization of quadratic forms.

In a discrete setting a positive-definite quadratic objective function, defined as
\be J = \Re\left(\frac{1}{2}\nvec{x}^*\nodot\mat{Q}\nodot\nvec{x} - \nvec{b}^*\nodot\nvec{x}\right)
\ee{optobj}
with arbitrary coefficients $\nvec{b}$ and $\mat{Q}$ (where $\mat{Q}$ can be assumed to be hermitian with no loss of generality), is minimized when its differential is zero:
\[\d J = \Re\left(\nvec{y}\nvec{\d x}\right)=0\quad \text{where}\quad \nvec{y}=\left(\mat{Q}\nodot\nvec{x} - \nvec{b}\right)^*.
\]

When the system is instantaneous, \ie\ responds with no time delay and its state doubles up as the variable to be optimized, the minimum of \eqref{optobj} can be attained by, either directly or iteratively, inverting the linear system $\mat{Q}\nodot\nvec{x} - \nvec{b}=0$. For the iterative solution of this optimization problem a number of classical \emph{gradient-based} iteration algorithms have been developed (gradient, conjugate gradient, BFGS, {etc.}, see \eg\ \cite{NumRec}) which rely on the sensitivity $\nvec{y}$ (another name of which is the \emph{gradient}) being computed at each iteration.

When the system is dynamic, one can in principle replace it by its resolvent and proceed in the same way, or exploit the fact that computing a sensitivity is exactly what adjoint equations are best at.
If, for instance, the state $\nvec{x}$ is indirectly determined by the discrete-time dynamical system of \S\ref{discrete-time} and $J$ is a quadratic form in the final state $\nvec{x}_N$, the objective \eqref{dtobj} becomes replaced by
\be\d J = \Re\left(\nvec{y}\nvec{\d x}_N\right)\quad \text{with}\quad \nvec{y}=\left(\mat{Q}\nodot\nvec{x}_N - \nvec{b}\right)^*.
\ee{varN}
The calculation of $\nvec{v}$ from this $\nvec{y}$ proceeds exactly as in \S\ref{discrete-time}, but a new $\nvec{y}$ must be considered every time $\nvec{x}_N$ changes. The result is what is often called \emph{direct--adjoint iteration}:

\begin{enumerate}
\item Start
\marginpar{\mptit{Direct--adjoint iteration}: to alternate solutions of the direct and adjoint problem in an iterative optimization loop.}
with a (possibly zero) guessed value for the initial condition $\nvec{x}_0$.
\item Perform a \emph{direct} solution of system \eqref{dtsystem} from time $0$ to time $N$.
\item Calculate $\nvec{y}$ from $\nvec{x}_N$ according to \eqref{varN}.
\item Perform an \emph{adjoint} solution of system \eqref{dtadjoint} from $\nvec{v}_{\!N}=\nvec{y}$ at time $N$ to get the sensitivity (the gradient) $\nvec{v}_0$ at time $0$.
\item Improve the value of $\nvec{x}_0$ using $\nvec{v}_0$ with one of the gradient-based optimization algorithms and loop back to step 2 until convergence.
\end{enumerate}

A similar direct--adjoint iteration can be applied to  optimize an objective depending on the state at all times of any of the discrete or continuous problem types considered in \S\ref{simplest}--\ref{spacepde}. Since the adjoint problem has the same computational cost as the direct problem, each iteration costs twice as much as a single direct solution with specified initial conditions and forcing.

By contrast, the alternate route of preliminarily calculating a resolvent matrix and then optimizing requires a number of direct-solution times equal to the dimension $M$ of $\nvec{x}_0$. The balance falls in favour of direct--adjoint iteration when the expected number of iterations to convergence is smaller than $M/2$.

\subsubsection{Optimal perturbation}
\label{optper}
In investigating the transient growth of boundary-layer disturbances
, a key role is played by the initial condition that maximizes the energy amplification of a system of the general form \eqref{dtsystem}, \ie\ an objective defined as $J=E_\text{out}/E_\text{in}$ where the input and output energies are quadratic forms $E_\text{in}=\nvec{x}_0^*\mat{Q}_\text{in}\nvec{x}_0$ and $E_\text{out}=\nvec{x}_N^*\mat{Q}_\text{out}\nvec{x}_N$ characterized by two, possibly different, hermitian matrices $\mat{Q}_\text{in}$ and $\mat{Q}_\text{out}$. The extremality condition for this Rayleigh ratio (or equivalently for $E_\text{out}$ under the constraint $E_\text{in}=1$) may be written in terms of the resolvent $\mat{H}$ as
\be \mat{H}^\mathrm{H}\mat{Q}_\text{out}\mat{H}\nvec{x}_0 = J\, \mat{Q}_\text{in}\nvec{x}_0
\ee{Rayleigh}
(where superscript $\,^\mathrm{H}$ denotes the hermitian transpose, \ie\ the complex-conju\-gate transpose). Since both $\mat{H}^\mathrm{H}\mat{Q}_\text{out}\mat{H}$ and $\mat{Q}_\text{in}$ are hermitian matrices, this can be recognized to be a self-adjoint eigen\-value problem with all the properties listed in \S\ref{selfadjoint}, in particular with all real eigen\-values of $J$. Therefore its maximum eigen\-value and corresponding optimal initial condition can be determined by a simple power iteration:
\be
\tilde{\nvec{x}}_0=\mat{Q}_\text{in}^{-1}\mat{H}^\mathrm{H}\mat{Q}_\text{out}\mat{H}\nvec{x}_0^{(i)};\quad
J=\sqrt{\tilde{\nvec{x}}_0^*\mat{Q}_\text{in}\tilde{\nvec{x}}_0};\quad   \nvec{x}_0^{(i+1)}=\tilde{\nvec{x}}_0/J
\ee{powerit}
(where superscript $^{(i)}$ counts iterations and $\tilde{\nvec{x}}_0$ is an intermediate variable). In practice, \eqref{powerit} can be implemented in the form of a direct--adjoint iteration without explicit use of the resolvent, if only it is recognized that multiplication by $\mat{H}$ is equivalent to one application of the direct and multiplication by $\mat{H}^\mathrm{H}$ to one application of the adjoint equations.

\stabilityref{Direct-adjoint iteration in the form \eqref{powerit} was instrumental to the computation of optimal inflow disturbances for a Blasius boundary layer by \cite{Andersson1bis,Andersson1} and \cite{Luchini2bis,Luchini2}, who achieved the quantitative calculation of non-parallel spatial transient growth in such a flow (a parallel approximation of which had been studied by \cite{Butler}). The more general gradient-based algorithm was used 
in the open-loop optimal control of linear \citep{Cathalifaud} and nonlinear \citep{Zuccher1} transient growth through blowing and suction at the wall.}

\subsubsection{Optimal control}
In another example, the optimization at hand might be a closed-loop optimal-control problem.
In the so-called \emph{linear-quadratic regulator} (LQR) setting (\eg\ \cite{Kim}), this is usually formulated as follows:

\marginpar{\mptit{LQR optimal control prob\-lem}: For every given initial condition find the input that minimizes a weighted sum of the state and control energies.}
\begin{verse}
\noindent For a dynamical system of equation \eqref{ctsystem}, and a given initial condition $\nvec{x}(0)=\nvec{x}_0$, find the input $\nvec{u}(t)$ that minimizes the quadratic form $J=\frac{1}{2}\int_0^T(\nvec{x}^*\mat{Q}\nvec{x}+\nvec{u}^*\mat{R}\nvec{u})\d t$
\end{verse}
(where $\mat{Q}$ and $\mat{R}$ can be assumed to be hermitian with no loss of generality). Considering the first variation of the objective,
\[\delta J\;=\int_0^T\Re\left(\nvec{x}^*\mat{Q}\nvec{\delta x}+\nvec{u}^*\mat{R}\nvec{\delta u}\right)\d t,
\]
and applying \eqref{ctadjoint}--\eqref{ctsolvedobj} with $\nvec{y}=\nvec{x}^*\mat{Q}$ yields the adjoint equation
\be-\,\frac{\d\nvec{v}}{\d t}=\mat{A}^{\!\trasp} \nvec{v}+\nvec{x}^*\mat{Q}.
\ee{controladjoint}
and the explicit expression of the objective variation
\[\delta J=-\Re\Big[\nvec{v}\nvec{\delta x}\Big]_0^T+
\int_0^T\Re\left[(\nvec{u}^*\mat{R}+ \nvec{v} \mat{B})\nodot \nvec{\delta u}\right] \d t.
\]
The variation $\delta J$ is then made to vanish by: 1) letting $\nvec{\delta x}(0)=0$ (since the initial condition $\nvec{x}(0)$ is imposed); 2) letting $\nvec{v}(T)=0$ (
the terminal condition for the backward integration of \eqref{controladjoint}); 3) letting
\be \nvec{u}= -\mat{R}^{-1} \mat{B}^\mathrm{H}\nvec{v}^*.
\ee{ucontrol}

At this stage
\marginpar{\mptit{Hamiltonian matrix}:\\
combining the direct and adjoint problem together, shares the same symmetry as the Hamilton equations of analytical mechanics.}
there are two possibilities. One is to write the system of equations \eqref{ctsystem} (with $\nvec{u}$ replaced by \eqref{ucontrol}) and \eqref{controladjoint} together as the $2M\times 2M$ linear problem
\be
\frac{d}{dt} \begin{bmatrix}
\nvec{x}\\
\overbar{\nvec{v}}
\end{bmatrix}
=
\begin{bmatrix}
\mat{A}&\mat{B}\mat{R}^{-1}\mat{B}^\mathrm{H}\\
\mat{Q}&-\mat{A}^\mathrm{H}
\end{bmatrix}
\begin{bmatrix}
\nvec{x}\\
\overbar{\nvec{v}}
\end{bmatrix}
\ee{hamiltonian}
where $\overbar{\nvec{v}}=-\nvec{v}^*$ and boundary conditions are $\nvec{x}(0)=\nvec{x}_0$, $\overbar{\nvec{v}}(T)=0$. This boundary-value problem, which shares the same symmetry as the Hamilton equations of analytical mechanics and is therefore also termed \emph{hamiltonian}, may then be dealt with \emph{as is} or transformed into the so-called \emph{matrix Riccati equation} \citep{Kim}. Its solution is classically obtained by applying the eigen\-vector expansion \eqref{eigenexpansion} of the resolvent, with the hamiltonian matrix in the place of $\mat{A}$. This is a standard procedure in control theory, for which easy to use computer libraries are available; however the resolvent matrix is not sparse, even when $\mat{A}$ is, and therefore its use may become impractical for very large state vectors such as obtained from the discretization of partial differential equations (\S\ref{spacepde}).

An alternative is to apply direct--adjoint iteration, by repeatedly solving \eqref{ctsystem} forward in time with a tentative $\nvec{u}$ and \eqref{controladjoint} backward in time for the sensitivity $\nvec{u}^*\mat{R}+ \nvec{v}\nodot \mat{B}$, using then one of the gradient-based optimization algorithms to produce an improved $\nvec{u}(t)$ for the next step \citep{Riccatiless}.

\subsubsection{Adjoint of the direct--adjoint}\label{ADA}
The full purpose of an optimal-control problem is not so much to solve the optimization problem \eqref{hamiltonian} for \emph{one} initial condition $\nvec{x}_0$, as to determine the matrix (\emph{feedback kernel}) $\mat{K}$ that provides the feedback control $\nvec{u}(0)$ for \emph{any} $\nvec{x}_0$ in the form of $\nvec{u}(0)=\mat{K}\nvec{x}_0$. It is sufficient for this purpose to obtain $\nvec{u}$ at time $t=0$, because $\nvec{u}$ for a later time will be calculated from the state at that time. That such a $\mat{K}$ exists is ensured by the problem \eqref{hamiltonian} being linear.

Notice that $\mat{K}$ is an $N\times M$ matrix, if $M$ is the dimension of the state and $N$ of the input. A possible direct way to determine $\mat{K}$ one column at a time is:
\begin{enumerate}
\item choose a basis of $M$ independent $M$-vectors for $\nvec{x}_0$ (\eg, the trivial basis $(1,0,0,\dots)$, $(0,1,0,\dots)$, \dots);
\item for each $\nvec{x}_0$ solve (either directly or iteratively) \eqref{hamiltonian} and extract the corresponding $\overbar{\nvec{v}}(0)$ (also an $M$-vector);
\item calculate $\nvec{u}(0)$ (an $N$-vector) from \eqref{ucontrol} and store it in the appropriate column of $\mat{K}$.
\end{enumerate}

The above procedure requires \eqref{hamiltonian} to be solved $M$ times. However if $N\ll M$ (a frequent occurrence), a much more efficient route exists: to calculate $\mat{K}$ one \emph{row} at a time by
applying adjoint analysis a second time over! Since \eqref{hamiltonian} is already a direct--adjoint problem, we name this approach ADA (\emph{adjoint of the direct--adjoint}) \citep{Riccatiless}.

At first thought it may seem like a strange idea to take the adjoint twice, but if the compounded hamiltonian problem \eqref{hamiltonian} is regarded as just another linear system with $\nvec{x}(0)$ as input and $\nvec{u}(0)$ as output, the way to proceed becomes straightforward. Contracting each side of the direct system \eqref{hamiltonian} with a new $2M$-vector which may be denoted $[{\nvec{x}^\dagger}^*,{\nvec{v}^\dagger}^*]$ and integrating by parts gives the adjoint system
\be
\frac{d}{dt} \begin{bmatrix}
\nvec{x}^\dagger\\
\nvec{v}^\dagger
\end{bmatrix}
=
\begin{bmatrix}
-\mat{A}^\mathrm{H}&-\mat{Q}\\
-\mat{B}\mat{R}^{-1}\mat{B}^\mathrm{H}&\mat{A}
\end{bmatrix}
\begin{bmatrix}
\nvec{x}^\dagger\\
\nvec{v}^\dagger
\end{bmatrix}
\ee{adjhamiltonian}
(account having been taken that both $\mat{Q}$ and $\mat{R}$ are hermitian) and boundary conditions ${\nvec{v}^\dagger}(0)=-\nvec{y}^*$, $\nvec{x}^\dagger(T)=0$, where we have assumed an objective of the form $J=\nvec{y}\overbar{\nvec{v}}(0)$. According to \eqref{ucontrol}, we can now identify this objective with one selected component of $\nvec{u}(0)$ by letting $-{\nvec{v}^\dagger}^*(0)$ equal the corresponding row of $\mat{R}^{-1}\mat{B}^\mathrm{H}$. The solution of the adjoint hamiltonian problem will then provide the sensitivity ${\nvec{x}^\dagger}^*(0)$ of this component of $\nvec{u}(0)$ to the entire initial condition $\nvec{x}(0)$, \ie\ one row of $\mat{K}$.

In fact, the process becomes even simpler because \eqref{adjhamiltonian} reduces easily to the same form as \eqref{hamiltonian}. (The hamiltonian problem is, in a suitable sense, self-adjoint, which is not so surprising because it already contains both direct and adjoint components.)
The two problems become identical, as can be easily verified, under the \emph{symplectic transformation} (well known from analytical mechanics, \eg\ \cite{HGoldstein} p. 343)
\[\nvec{v}^{\dagger}=-\nvec{x},\quad \nvec{x}^\dagger=\overbar{\nvec{v}}.
\]
Therefore, whatever algorithm is already available to solve the hamiltonian problem \eqref{hamiltonian} it may be reused for its adjoint, by just assigning $\nvec{y}^*$ as the initial condition for $\nvec{x}(0)$ and taking $\overbar{\nvec{v}}^*(0)$ as the result.

In conclusion we arrive at the following ADA procedure:
\marginpar{\mptit{ADA}: adjoint of the direct--adjoint hamiltonian system.}
\begin{enumerate}
\item choose a basis of $N$ independent $M$-vectors for the row space of $\mat{R}^{-1}\mat{B}^\mathrm{H}$ (in practice, take each row of this matrix in turn);
\item using each $M$-vector as the initial condition $\nvec{x}^*(0)$, solve (either directly or iteratively) \eqref{hamiltonian} and extract the corresponding $\overbar{\nvec{v}}(0)$;
\item assign $\overbar{\nvec{v}}^*(0)$ (also an $M$-vector) to the corresponding row of $\mat{K}$.
\end{enumerate}
This algorithm achieves to calculate $\mat{K}$ by rows, rather then by columns, using $N$ (the dimension of the input) rather than $M$ (the dimension of the state) independent solutions of \eqref{hamiltonian}. 

\stabilityref{
The ADA algorithm is part of an ongoing effort to devise Riccati-less procedures to solve control problems, \ie\ iterative algorithms that can exploit 
sparseness and do not require the solution of a matrix Riccati equation. While, in fact, the standard approach is adequate for problems with, say, a 
thousand state variables, the discretization of a fluid flow problem 
easily contains millions or hundreds of millions of variables, 
and a full matrix of a million by a million elements is not a viable tool. The frequently adopted route of model reduction (replacing the original problem by a model having 
a smaller number of state variables) in a sense defeats the purpose of optimization, unless the model itself is optimized which is a task of comparable 
difficulty to optimizing the controller. On the other hand, an iterative algorithm can deal with such problems without any model reduction. 
An overview of several suitable iterative approaches (one of which is ADA) is given by \cite{Bewley}.
\marginpar{\mptit{AAD} Adjoint of the adjoint-direct hamiltonian problem.}
 The dual version of ADA for an estimation problem
 (every control problem has a dual estimation problem, see \eg\ \cite{Kim}) is described under the acronym of AAD by \cite{Semeraro}. 
}

\subsection{The space continuum: adjoint of a partial differential equation}\label{spacepde}
The procedure outlined in \S\ref{contadj}--\ref{bvpevp} generalizes unchanged to linear partial differential equations involving a function $f$ of time (which may be renamed $x_0$) and $H$ other (typically spatial) independent variables $x_1\ldots x_H$. Formally, the equation (or system of equations) is multiplied by an as yet unknown function (or vector of such functions) $v$ and integrated over the solution domain $\Omega$. Subsequently each term involving a derivative of $f$ is integrated by parts one or more times until the original unknown $f$ can be factored out and only derivatives of $v$ appear.

The multidimensional equivalent of integration by parts is known as the Lagrange-Green identity
(\eg, \cite{FJohn} p. 79). For a single term of the form $a\,\deh{b}{x_h}$ this identity (a consequence of Gauss' theorem applied to the integral of $\deh{(a\,b)}{x_h}$) may be written as
\marginpar{\mptit{Lagrange--Green} identity.}
\be\underbrace{\int_\Omega a\,\de{b}{x_h}\d \Omega}_1\, = \underbrace{\oint_{\partial\Omega} a\,b\, n_h \d\partial \Omega}_2\,  \underbrace{-\int_\Omega \de{a}{x_h}\,b \d \Omega}_3.
\ee{type3}
where $n_h$ is the $h$-th component of the outward normal versor $\nvec{n}$. Repeatedly applying this transformation as many times as needed to each and every derivative in the partial differential equation
\be\op{L}(f)=u
\ee{pardiffeq}
(where $\op{L}$ is a differential operator representing a linear combination of partial derivatives of its argument) leads to an identity which can be formally written as
\be\int_\Omega v \op{L} (f) \d\Omega \,= \oint_{\partial\Omega} \op{M}(v,f,\nvec{n})\d \partial\Omega \,+ \int_\Omega \op{L}^\dagger (v) f \d\Omega
\ee{LagrangeGreen}
where the boundary integral collects all terms denoted as type 2 in \eqref{type3} ($\op{M}$ being linear in each of its three arguments) and $\op{L}^\dagger$ (also a differential operator representing, like $\op{L}$, a linear combination of partial derivatives of its argument) all terms of type 3.

Finally, replacing $\op{L}(f)$ by $u$ and equating the last term of \eqref{LagrangeGreen} to the objective
\be J\;=\int_\Omega y f \d \Omega
\ee{pdeobj}
yields the adjoint differential equation
\be\op{L}^\dagger (v) = y
\ee{pdeadjoint}
and the explicit expression of the objective as a function of the problem's \rhs\ and boundary data:
\be J\;=\int_\Omega v u \d\Omega - \oint_{\partial\Omega} \op{M}(v,f,\nvec{n})\d \partial\Omega.
\ee{pdesolvedobj}
 
The only peculiar difficulty (which is sometimes turned into an advantage) 
arises in specifying suitable functional spaces to which $f$ and $v$ must belong in order to ensure that all 
intermediate expressions exist.
Algebra defines the concept of \emph{dual} space
\marginpar{\mptit{Dual space}: vector space of all linear forms over a given space.}
as the vector space of all linear forms over a given vector space. 
Therefore, \emph{a priori} $f$ must belong to a (typically Sobolev) space regular enough that all required derivatives exist and $v$ to its dual 
space.
This is referred to as the \emph{strong formulation} of the differential problem. 
However, after the integration by parts only derivatives of $v$ appear; $v$ is then required to be sufficiently regular whereas 
regularity requirements on $f$ can be relaxed. The transformation to adjoint form allows the differential problem to be 
defined also for functions which are not as highly differentiable, and provides what is called its \emph{weak formulation}.

\subsubsection{The example of Green's formula}
Probably the most classical example of adjoint of a partial differential
\marginpar{\mptit{George Green}'s privately published comprehensive essay

\centering
\href{http://books.google.it/books?id=GwYXAAAAYAAJ}{\includegraphics[width=\marginparwidth,clip=true,trim=0.65cm 0cm 0.55cm 0cm]{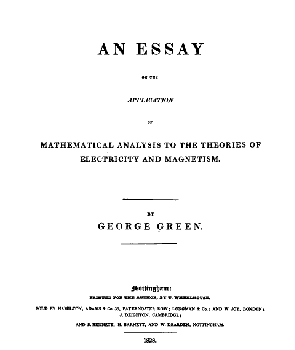}}}
equation is Green's formula for the Poisson equation \citep{Green}, together with the related reciprocity theorems for the wave equation and electromagnetic fields.

Starting from the two-dimensional Poisson equation as an example, namely
\be
\de{^2 f}{x_1^2} + \de{^2 f}{x_2^2} = u ,
\ee{Poisson}
we multiply both sides
by $v$ and integrate. Then \eqref{type3} is applied twice to the first and 
twice to the second term. There results the familiar Green's identity that is usually expressed through the $\bnabla$ operator as
\be\int_\Omega v\nabla^2 f  \d \Omega \,= \oint_{\partial\Omega} \left(v\de{f}{n}-f\de{v}{n}\right)\d\partial\Omega  \,+\int_\Omega f \nabla^2 v  \d \Omega.
\ee{Greenid}
Defining an objective $J$ according to \eqref{pdeobj} and comparing \eqref{Greenid} with \eqref{LagrangeGreen} then yields the explicit expression of the objective
\[
J\;=\int_\Omega v u \d\Omega-\oint_{\partial\Omega} \left(v\de{f}{n}-f\de{v}{n}\right)\d \partial\Omega
\]
and the adjoint differential equation
\[\nabla^2 v = y.
\]
This is again a Poisson equation formally similar to \eqref{Poisson}, as originally discovered by Green, a fact which is today denoted by qualifying the Poisson problem as self-adjoint.
The availability of an adjoint (be it coincident or not with the direct differential equation) implies that the objective $J$ can be calculated multiple times as an integral over the direct problem's forcing and boundary data after \eqref{pdeadjoint} has been solved just once. For the Poisson equation, Green's result is most frequently applied with $y$ equal to a Dirac delta function (\eg\ \cite{Greenberg}) so that the objective becomes the value of $f$ at a target point, but the potentiality of the adjoint also remains open for any other objective and its corresponding $y$.

\subsection{Adjoint of a nonlinear problem}\label{nonlinear}
As was remarked in \S\ref{simplest} the sensitivity, or the adjoint, we have
\marginpar{\mptit{Sensitivity}: derivative of\linebreak the objective function with respect to the forcing and boundary conditions.}
until now spoken of is defined as the derivative of the objective with respect to the input data of the problem. While for a linear problem this derivative is constant and can be related to the transpose of a matrix, for a more general nonlinear problem the adjoint will depend on the state itself and lose its linear-algebra interpretation. Nonetheless it will continue to make perfect sense as a derivative, something we have already made use of in \S\ref{optimization}.

The nonlinear generalization of the discrete-time system \eqref{dtsystem} (which includes \eqref{gdtsystem} as remarked at the end of \S\ref{discrete-time}) can be written as a state-transition equation of the form
\( \nvec{x}_{n+1}=\nvec{f}_n(\nvec{x}_n)\) (where $n=0\ldots N-1$),
with initial condition $\nvec{x}_0$ and objective
\( J=F(\nvec{x}_N).
\) 
The derivative of $J$ with respect to the initial condition is given by the chain rule of differentiation as
\[\frac{\d J}{\d \nvec{x}_0} = \de{J}{\nvec{x}_N}\nodot\de{\nvec{x}_N}{\nvec{x}_{N-1}}\nodot \cdots\de{\nvec{x}_2}{\nvec{x}_{1}}\nodot\de{\nvec{x}_1}{\nvec{x}_{0}} ,
\]
where each $\deh{\nvec{x}_{n+1}}{\nvec{x}_n}$ is a square jacobian matrix. In close analogy to \eqref{dtadjoint}, this repeated product may be associated to the left and written as a backwards recursion:
\be
\nvec{v}_{\!n}=\left(\de{\nvec{x}_{n+1}}{\nvec{x}_{n}}\right)^\trasp\nodot\nvec{v}_{\!n+1}\qquad(n=N-1\ldots 0),
\ee{nldtadjoint}
with terminal condition $\nvec{v}_{\!N}=\partial J/\partial \nvec{x}_N$ and eventual result $\nvec{v}_0=\d J/\d \nvec{x}_0$. Since each $\nvec{v}_{\!n}$ is a vector, not a matrix, \eqref{nldtadjoint} is no more computationally cumbersome than \eqref{dtadjoint} and can exploit sparseness; the same remains true for other discrete or continuous problem variations.

In many practical applications, the transition function $\nvec{f}_n$ will be such that only a few components of the state $\nvec{x}$ actually change in each step. The direct and adjoint systems may then be usefully rewritten in the form
\marginpar{\mptit{Nonlinear} discrete-time dynamical system and its adjoint.}
\be \nvec{x}_{n+1}=\nvec{x}_{n}+\nvec{g}_n(\nvec{x}_n)\qquad(n=0\ldots N-1)
\ee{inldtsystem}
\be
\nvec{v}_{\!n}=\nvec{v}_{\!n+1}+\left(\de{\nvec{g}_{n}}{\nvec{x}_{n}}\right)^\trasp\nodot\nvec{v}_{\!n+1}\qquad(n=N-1\ldots 0),
\ee{inldtadjoint}
where $\nvec{f}_n= \nvec{x}_n+\nvec{g}_n$ and just the nonzero components of $\nvec{g}_{n}$ and $\deh{\nvec{g}_{n}}{\nvec{x}_{n}}$ need be acted upon. Notice that the adjoint vector $\nvec{v}$ is sparsely altered, not in those components whose state changes in each step, but in those which the state components that change depend upon.

The only significant computational aggravation of the nonlinear adjoint is in its memory requirements: $\partial{\nvec{g}_{n}}/\partial{\nvec{x}_{n}}$ is a function of $\nvec{x}_{n}$, and since the loop in \eqref{inldtadjoint} runs opposite to the loop in \eqref{inldtsystem}, the state $\nvec{x}_{n}$ has to be permanently stored (or at least \emph{checkpointed}, see \cite{Griewank}) for it to be accessible in reverse order.

\subsection{Adjoint of the Navier--Stokes problem}
\label{NavierStokes}
The most general fluid-mechanics problem we shall consider in this review is one governed by the Navier--Stokes equations for an incompressible fluid with constant viscosity:
\begin{subequations}
\begin{gather}
  \dfrac{\partial \vec{v}}{\partial t}+\nabla\cdot\left(\vec{v}\otimes \vec{v}\right)+\nabla p =\dfrac{1}{\Rey}\nabla^2 \vec{v}+\vec{f},
 \\
\nabla\cdot \vec{v}=\dot{m}.
\end{gather}
\label{NS}
\end{subequations}
Here $\vec{v}$ is the dimensionless velocity vector\footnote{The $\vec{\ }$ modifier is used to denote vectors in physical space, as distinguished from numerical vectors of arbitrary dimension%
. The scalar product of two space vectors is explicitly indicated by a dot, $\nabla\cdot$ denotes the divergence operator
.}, $p$ is the reduced pressure, $\Rey$ is the Reynolds number and $\vec{f}$, $\dot{m}$ are an external volume force and mass source\footnote{Notice that in the presence of a mass source the conservative and convective forms of the momentum equation are no longer equivalent.} which will serve the role of the \rhs\ $u$ in \eqref{pardiffeq} or \eqref{Poisson}. No obstacle prevents us
from applying adjoint techniques to compressible, non-newtonian or even more general fluids, but for definiteness we shall limit ourselves to \eqref{NS}.

Equations \eqref{NS} can be linearized by letting
$\vec{v}\leftarrow \vec{v}+\vec{\delta v}$ and similarly for the other variables, where symbols prefixed by a $\delta$ are infinitesimal in the sense of variational calculus. This results in the linearized Navier--Stokes equations (LNSE):
\marginpar{\mptit{LNSE}: Linearized Navier\--Stokes equations}
\begin{subequations}
\begin{gather}
\de{\vec{\delta v}}{t}+ \nabla\cdot\left(\vec{v}\otimes\vec{\delta v}+\vec{\delta v}\otimes\vec{v}\right)+\nabla \delta p-\dfrac{1}{\Rey}\nabla^2\vec{\delta v}=\vec{\delta f} , \label{LNSEa} \\
 \nabla\cdot \vec{\delta v}=\delta \dot{m} . \label{LNSEb}
\end{gather}
\label{LNSE}
\end{subequations}

An arbitrary, generally nonlinear, functional $J$ of the velocity and pressure fields may be introduced as the objective. Examples of such functionals are the integral of the kinetic-energy density $v^2$ over the volume $\Omega$ or part of it (possibly the inlet or outlet surface), the lift or drag (momentum flux through the solid boundaries), {etc.} The first variation $\delta J$ of the objective functional will be a linear functional of $\vec{\delta v}$ and $\delta p$, expressible in the integral form
\[\delta J\; = \int_0^T\int_\Omega\left(\op{D}_{\vec{v}} J\cdot \vec{\delta v} + \op{D}_p J\;\delta p\right) \d\Omega \d t
\]
where $\op{D}_{\vec{v}} J$ and $\op{D}_p J$ (possibly distributions) are named \emph{Fr\'echet derivatives} of $J$ with respect to velocity and pressure.

In order to apply the Lagrange-Green identity \eqref{LagrangeGreen}, we can multiply \eqref{LNSEa} and \eqref{LNSEb} by two as yet unspecified functions  (whose meaning will become clear later) $\vec{f}^{\,\dagger}$ and $\dot{m}^\dagger$, and integrate by parts as many times as needed to let derivatives of $\vec{\delta v}$ and $\delta p$ disappear. Doing so yields the adjoint equations
\marginpar{\mptit{Adjoint Navier--Stokes}\\ equations.}
\begin{subequations}
\begin{gather}
\de{\vec{f}^{\,\dagger}}{t}+\vec{v}\cdot\left( \nabla \vec{f}^{\,\dagger}+{\nabla \vec{f}^{\,\dagger}}^\trasp\right)+\nabla \dot{m}^\dagger +\dfrac{1}{\Rey}\nabla^2\vec{f}^{\,\dagger}=-\op{D}_{\vec{v}} J , \\
\nabla\cdot\vec{f}^{\,\dagger}=-\op{D}_p J ,
\end{gather}
\label{adjNS}
\end{subequations}
plus the explicit expression of the objective variation as a function of the forcing and boundary conditions:
\begin{multline}
\label{LGNS}
\delta J =  \int_0^T\int_\Omega\left( \vec{f}^{\,\dagger}\cdot\vec{\delta f} +  \dot{m}^\dagger \delta\dot{m}\right) \d \Omega \d t - \left[\int_\Omega \vec{f}^{\,\dagger}\cdot\vec{\delta v} \d \Omega \right]_0^T + \null \\
- \int_0^T\oint_{\partial\Omega} \left[\left(\vec{f}^{\,\dagger}\cdot\vec{\delta v}\right)\,\vec{v}+ \left(\vec{f}^{\,\dagger}\cdot\vec{v}+\dot{m}^\dagger\right)\,\vec{\delta v}+\null\right.\\
\left.\null+\dfrac{1}{\Rey} \left(\nabla \vec{f}^{\,\dagger}\cdot \vec{\delta v}-\nabla \vec{\delta v} \cdot \vec{f}^{\,\dagger}  \right)+\vec{f}^{\,\dagger}\;\delta p \right]\cdot\vec{n}\d \partial\Omega \d t.
\end{multline}

Just as in \S\ref{bvpevp}, the boundary conditions for \eqref{adjNS} must be chosen by the rule that, term by term and portion by portion of the boundary, \eqref{LGNS} should only depend on the physically assigned boundary conditions for \eqref{LNSE} (for instance, $\vec{f}^{\,\dagger}$ must be zero at final time if $\vec{\delta v}$ is assigned at initial time, or both must be periodic if an oscillating solution is sought). It then becomes apparent from the first term of \eqref{LGNS} that the adjoint
\marginpar{\mptit{Variables of the adjoint Navier--Stokes problem}: Fr\'echet derivatives of the objective with respect to volume force and mass injection rate.}
variables $\vec{f}^{\,\dagger}$ and $\dot{m}^\dagger$ have the meaning of Fr\'echet derivatives of $J$ with respect to volume force and mass injection 
rate, while the remaining terms provide the sensitivities of $J$ to the boundary conditions, which are of importance in 
the \emph{receptivity} analysis of boundary layers
. In a problem with no volume force or mass injection, only the sensitivities at the boundaries 
are used but the meaning of $\vec{f}^{\,\dagger}$ and $\dot{m}^\dagger$ remains unchanged, just as in Green's formula for the Laplace rather than the Poisson equation.

\subsection{Writing an adjoint computer program: adjoint of the discretization versus discretization of the adjoint}
\label{AdjointProgram}
As clearly appears from the considerations at the beginning of \S\ref{contadj}, there is a complete correspondence between the adjoint of an integral or differential problem and the adjoint of a discrete approximation of the same, provided the approximation is convergent. Therefore in numerical applications a choice can be operated between writing an adjoint differential equation such as \eqref{pdeadjoint} or \eqref{adjNS} and then discretizing it
\marginpar{\mptit{DA}: Discretization of the Adjoint.

\mptit{AD}: Adjoint of the Discretization (also used for Automatic Differentiation, with essentially the same meaning).
}
(discretization of the adjoint, DA) or starting from a discretization of the direct problem in the form \eqref{isdsystem} and obtaining its adjoint (adjoint of the discretization, AD) as in \eqref{isdadj}. For most purposes the choice is a matter of personal preference
, however a few subtle points are involved:
\begin{enumerate}
\item Although DA and AD converge to the same result (when they converge at all, which case by case may or may not be an easy proposition to prove mathematically), they do not necessarily share the same discretization error for nonzero step size.
\item Convergence may be in a weak (integral) sense rather than pointwise, especially when derivatives are involved; the evaluation of the sensitivity to boundary conditions requires special care.
\item For efficiency the AD approach should not be taken literally to mean that an overall resolvent matrix must be built and then transposed, but rather that each intermediate stage of the numerical procedure must be taken an adjoint of. Fortunately the process is fairly straightforward, and can even be automated through automatic differentiation software \citep{Griewank}. 
\end{enumerate}

Starting from the last point, every computer program is a dynamical system of the form \eqref{inldtsystem}, where $n$ numbers the individual steps of program execution (\ie\ program statements or blocks of statements in the order they are executed, inclusive of the effect of any branches and loops, as opposed to the order in which they are written) and $\nvec{x}_n$ is the set of all variables the program keeps in memory
at the beginning of each step. The key to, either manually or automatically, writing an AD program is to run through the same sequence of steps backwards, each time updating by \eqref{inldtadjoint} the sensitivity of those variables only on which the present step depends. The result will be an adjoint computer program that has a comparable computation time to (exactly the same number of steps as) the direct program.

Convergence of AD is not an issue when the optimal value of an objective is being determined. In this case, in fact, the sensitivity must be driven to zero, and doing so will provide the exact optimum of the discretized objective. If the direct discretization makes any sense, this is the desired result.

When, on the other hand, the sensitivity itself is the target of the computation, the choice between AD and DA is less clear-cut. An example may serve to clarify the potential pitfalls involved.

If the integral \eqref{ctobj} defines the objective of an instantaneous system (whose state $\nvec{x}$ coincides for simplicity with its input $\nvec{u}$), the sensitivity of $J$ to $\nvec{x}$ is, trivially, $\nvec{y}$. If now this integral is discretized by a first-order rectangle rule,
\[J_{\text{r}} = \Delta t\sum_{n=0}^{N-1} \nvec{y}(n\Delta t)\nvec{x}(n\Delta t),
\]
the derivative of $J_{\text{r}}$ with respect to $\nvec{x}(n\Delta t)$ is $\nvec{y}(n\Delta t)\Delta t$, proportional to a first-order 
discretization of the exact result as may be expected (but notice that when the discretization is not uniform the factor $\Delta t$ must be handled properly).  In order to obtain a higher precision, one may decide to discretize the integral by 
the composite Simpson rule,
\[J_{\text{S}} = \Delta t\sum_{n=0}^{N} w_n \nvec{y}(n\Delta t)\nvec{x}(n\Delta t),
\]
\marginpar{\mptit{Composite Simpson rule}\\ and its adjoint.}
where the weights $w_n$ equal $1/3$ for $n=0$, $n=N$ and otherwise $4/3$ for $n$ odd, $2/3$ for $n$ even.

The derivative of $J_{\text{S}}$ with respect to $\nvec{x}(n\Delta t)$ will then turn out to be\linebreak
$w_n\nvec{y}(n\Delta t)\Delta t$, a function oscillating from one grid point to the next which does not converge at all (in a pointwise sense) to the desired $\nvec{y}(n\Delta t)$. On the other hand, the integral of the product between $\nvec{y}$ and any smooth function of compact support converges with $4^\text{th}$-order or better accuracy, which shows that convergence can be recovered by a suitable smoothing (or by adopting a non-oscillatory quadrature formula, \eg\ \cite{Luchiniquadr}).

In quite a few problems of mathematical physics, among which the Navier--Stokes equations \eqref{NS}, it so occurs that the highest-derivative 
terms form a self-adjoint system (the steady Stokes problem is self-adjoint under the transformation $\vec{f}^{\,\dagger}=\vec{v}$, $-\dot{m}^\dagger=p$, 
as may easily be verified by dropping the convective terms in \eqref{NS} and \eqref{adjNS}, a fact which probably inspired \cite{Hill1} and other authors 
to adopt the names of \emph{adjoint velocity} and \emph{adjoint pressure} for these quantities). In this case AD and DA have equivalent smoothing properties: 
paying attention to choose a self-adjoint discretization one can get the best of both worlds. For instance, many finite-element discretizations are self-adjoint by construction (they were originally designed for problems that admit a variational principle, which is another facet of being self-adjoint). Finite differences are frequently self-adjoint on a uniformly spaced grid but may require special care on non-uniform ones. Essentially non-self-adjoint, on the other hand, are problems involving inviscid flow, especially in the presence of shocks; these are typical cases where the difference between AD and DA
must be given the greatest consideration \citep{Jameson2}. \pagebreak

An easy test to validate
\marginpar{\mptit{Validating an adjoint pro\-gram}: run the direct program with a random $\nvec{u}$, the adjoint program 
with a random $\nvec{y}$, and then compare the values of $\nvec{y}\nodot\nvec{x}$ and $\nvec{v}\nodot\nvec{u}$.}
an adjoint program (or part of it) is to verify that the defining identity between \eqref{sdobj} and \eqref{sdobjsolved} be satisfied 
for two randomly generated vectors $\nvec{u}$ and $\nvec{y}$. This test just requires running the direct program with $\nvec{u}$ as data, the adjoint program 
with $\nvec{y}$, and then comparing the two linear forms $\nvec{y}\nodot\nvec{x}$ and $\nvec{v}\nodot\nvec{u}$.
When AD is adopted, the values of $J$ obtained must match within machine accuracy independently of the discretization stepsize, which is very unlikely to happen by chance in the presence of programming errors. Even more forcefully \eqref{fundid} must be satisfied at every intermediate step, which allows a faulty step to be quickly identified. The availability of this test alone is, in our experience, sufficient motivation to favour AD over DA whenever applicable.

\bibliography{adjoints,biblioFlavio}

\begin{thebibliography}{}
\expandafter\ifx\csname natexlab\endcsname\relax\def\natexlab#1{#1}\fi

\bibitem[{Airiau et~al.(2002)Airiau, Walther \& Bottaro}]{Airiau2}
Airiau C, Walther S, Bottaro A. 2002.
\newblock Boundary layer sensitivity and receptivity.
\newblock \textit{{C. R. M\'ecanique}} 330:259--265

\bibitem[{Andersson et~al.(1998)Andersson, Berggren \&
  Henningson}]{Andersson1bis}
Andersson P, Berggren M, Henningson DS. 1998.
\newblock Optimal disturbances in boundary layers.
\newblock In \textit{{Proc. AFOSR Workshop on Optimal Design and Control,
  Arlington VA, USA, 30 Sept--3 Oct 1997}}, eds. JT~Borggaard, J~Burns, E~Cli,
  S~Schreck. Birkhauser

\bibitem[{Andersson et~al.(1999)Andersson, Berggren \& Henningson}]{Andersson1}
Andersson P, Berggren M, Henningson DS. 1999.
\newblock Optimal disturbances and bypass transition in boundary layers.
\newblock \textit{Phys. Fluids} 11:134--150

\bibitem[{Bewley et~al.(2013)Bewley, Luchini \& Pralits}]{Bewley}
Bewley TP, Luchini P, Pralits J. 2013.
\newblock \bibmarginpar{New adjoint-based algorithms for the efficient solution
  of LQR problems, including the ADA approach}{Methods for the solution of
  large optimal control problems that bypass open-loop model reduction}.
\newblock \textit{IEEE Trans. Autom. Control} {s}ubmitted, available at
  \url{http://fccr.ucsd.edu/pubs/BLP_MCE_ADA_OSSI.pdf}

\bibitem[{Bottaro et~al.(2003)Bottaro, Corbett \& Luchini}]{Bottaro2}
Bottaro A, Corbett P, Luchini P. 2003.
\newblock \bibmarginpar{Adjoint-based optimisation of base-flow defects;
  introduction of structured pseudospectrum}{The effect of base flow variation
  on flow stability}.
\newblock \textit{J. Fluid Mech.} 476:293--302

\bibitem[{Butler \& Farrell(1992)}]{Butler}
Butler KM, Farrell BF. 1992.
\newblock Three-dimensional optimal perturbations in viscous flows.
\newblock \textit{Phys. Fluids A} 4:1637--1650

\bibitem[{Cathalifaud \& Luchini(2000)}]{Cathalifaud}
Cathalifaud P, Luchini P. 2000.
\newblock Algebraic growth in boundary layers: optimal control by blowing and
  suction at the wall.
\newblock \textit{Europ. J. Mech. B/Fluids} 19:469--490

\bibitem[{Courant \& Hilbert(1924)}]{Courant}
Courant R, Hilbert D. 1924.
\newblock \textit{Methoden des mathematischen {P}hysik, I}.
\newblock Verlag von Julius Springer

\bibitem[{Craig(1889)}]{Craig}
Craig T. 1889.
\newblock \textit{A treatise on linear differential equations}.
\newblock John Wiley \& Sons

\bibitem[{Dunford \& Schwartz(1958)}]{Dunford}
Dunford N, Schwartz JT. 1958.
\newblock \textit{Linear Operators part I: General Theory}.
\newblock Interscience

\bibitem[{Fedorov(1984)}]{Fedorov}
Fedorov AV. 1984.
\newblock Excitation of {T}ollmien-{S}chlichting waves in a boundary layer by a
  periodic external source located on the body surface.
\newblock \textit{Fluid Dyn.} 19:888--893

\bibitem[{Forsyth(1888)}]{Forsyth}
Forsyth AR. 1888.
\newblock Invariants, covariants, and quotient derivatives associated with
  linear differential equations.
\newblock \textit{Phil. Trans. R. Soc.} 179:377--489

\bibitem[{Fuchs(1873)}]{Fuchs}
Fuchs L. 1873.
\newblock {\"U}ber relationen, welche f{\"u}r die zwischen je zwei
  singul{\"a}ren {P}unkten erstreckten {I}ntegrale der {L\"o}sungen linearer
  {D}ifferentialgleichungen stattfinden.
\newblock \textit{Journal f{\"u}r die reine angewandte Mathematik} 76:177--213

\bibitem[{Giannetti \& Luchini(2003)}]{Giannetti03}
Giannetti F, Luchini P. 2003.
\newblock Receptivity of the circular cylinder's first instability.
\newblock In \textit{The 5th Euromech Fluid Mechanics Conference, Toulouse,
  France, August 24-28, 2003}. IMFT

\bibitem[{Giannetti \& Luchini(2007)}]{Giannetti1}
Giannetti F, Luchini P. 2007.
\newblock \bibmarginpar{Introduction of structural sensitivity
  analysis}{Structural sensitivity of the first instability of the cylinder
  wake}.
\newblock \textit{J. Fluid Mech.} 581:167--197

\bibitem[{Goldstein(1980)}]{HGoldstein}
Goldstein H. 1980.
\newblock \textit{Classical Mechanics}.
\newblock Addison--Wesley

\bibitem[{Golub \& {Van Loan}(1983)}]{Golub}
Golub GH, {Van Loan} CF. 1983.
\newblock \textit{Matrix Computations}.
\newblock Johns Hopkins University Press

\bibitem[{Green(1828)}]{Green}
Green G. 1828.
\newblock \textit{An essay on the application of mathematical analysis to the
  theories of electricity and magnetism}.
\newblock Printed for the {author} by T. Wheelhouse

\bibitem[{Greenberg(1988)}]{Greenberg}
Greenberg MD. 1988.
\newblock \textit{Advanced engineering mathematics}.
\newblock Prentice-Hall

\bibitem[{Griewank \& Walther(2008)}]{Griewank}
Griewank A, Walther A. 2008.
\newblock \textit{Evaluating Derivatives. Principles and Techniques of
  Algorithmic Differentiation}.
\newblock SIAM

\bibitem[{Hill(1992)}]{Hill1}
Hill DC. 1992.
\newblock \bibmarginpar{First adjoint-based global mode stabilization}{A
  theoretical approach for analyzing the restabilization of wakes}.
\newblock In \textit{AIAA paper 1992-0067}

\bibitem[{Hill(1995)}]{Hill2}
Hill DC. 1995.
\newblock \bibmarginpar{Local, adjoint-based receptivity analysis}{Adjoint
  systems and their role in the receptivity problem for boundary layers}.
\newblock \textit{J. Fluid Mech.} 292:183--204

\bibitem[{Jameson(2003)}]{Jameson2}
Jameson A. 2003.
\newblock Aerodynamic shape optimization using the adjoint method.
\newblock In \textit{Lecture Notes, von Karman Institute for Fluid Dynamics,
  Brussels, Belgium, Feb. 2003}

\bibitem[{John(1982)}]{FJohn}
John F. 1982.
\newblock \textit{Partial Differential Equations}.
\newblock Springer--Verlag

\bibitem[{Kim \& Bewley(2007)}]{Kim}
Kim J, Bewley TR. 2007.
\newblock A linear systems approach to flow control.
\newblock \textit{Annu. Rev. Fluid Mech.} 39:383--417

\bibitem[{Lagrange(1763)}]{Lagrange}
Lagrange JL. 1763.
\newblock Solution de diff\'erens {probl\^emes} de calcul int\'egral.
\newblock \textit{Miscellanea Taurinensia} 3:179--380.
\newblock [also in Oeuvres de Lagrange, J.A. Serret, ed., Vol. 1, 1867, pp.
  471-668]

\bibitem[{Luchini(1994)}]{Luchiniquadr}
Luchini P. 1994.
\newblock End-correction integration formulae with optimized terminal sampling
  points.
\newblock \textit{Comput. Phys. Commun.} 83:236--244

\bibitem[{Luchini(1997)}]{Luchini2bis}
Luchini P. 1997.
\newblock Effects on a flat--plate boundary layer of free--stream longitudinal
  vortices: optimal perturbations.
\newblock In \textit{EUROMECH 3rd European Fluid Mechanics Conference,
  Göttingen 15-18 Sept. 1997}. DLR

\bibitem[{Luchini(2000)}]{Luchini2}
Luchini P. 2000.
\newblock Reynolds-number-independent instability of the boundary layer over a
  flat surface: Optimal perturbations.
\newblock \textit{J. Fluid Mech.} 404:289--309

\bibitem[{Luchini \& Bottaro(1996)}]{Luchini1bis}
Luchini P, Bottaro A. 1996.
\newblock A time-reversed approach to the study of {G\"ortler} instabilities.
\newblock In \textit{Advances in Turbulence VI, Proc. 6th Europea n Turbulence
  Conference, Lausanne 2-5 July 1996}, eds. S~Gavrilakis, L~Machiels,
  PA~Monkewitz. Kluwer

\bibitem[{Luchini \& Bottaro(1998)}]{Luchini1}
Luchini P, Bottaro A. 1998.
\newblock \bibmarginpar{First non-local adjoint analysis for
  receptivity}{G\"{o}rtler vortices: a backward-in-time approach to the
  receptivity problem}.
\newblock \textit{J. Fluid Mech.} 363:1--23

\bibitem[{Pralits \& Luchini(2010)}]{Riccatiless}
Pralits JO, Luchini P. 2010.
\newblock Riccati-less optimal control of bluff-body wakes.
\newblock In \textit{Proc. 7th IUTAM Symposium on Laminar-Turbulent Transition,
  June 23-26, 2009, Stockholm, Sweden.}, eds. P~Schlatter, DS~Henningson.
  Springer

\bibitem[{Press et~al.(1999)Press, A.Teukolsky, Vetterling \&
  Flannery}]{NumRec}
Press WH, A.Teukolsky S, Vetterling WT, Flannery BP. 1999.
\newblock \textit{Numerical Recipes}.
\newblock Cambridge University Press

\bibitem[{Semeraro et~al.(2013)Semeraro, Pralits, Rowley \&
  Henningson}]{Semeraro}
Semeraro O, Pralits JO, Rowley CW, Henningson DS. 2013.
\newblock \bibmarginpar{New adjoint-based algorithms for estimation, including
  the AAD approach}{Riccati-less approach for optimal control and estimation:
  an application in {2D} boundary layers}.
\newblock \textit{J. Fluid Mech.} {s}ubmitted

\bibitem[{Tricomi(1957)}]{Tricomi}
Tricomi FG. 1957.
\newblock \textit{Integral Equations}.
\newblock Interscience

\bibitem[{Tumin \& Fedorov(1984)}]{Tumin3}
Tumin AM, Fedorov AV. 1984.
\newblock Instability wave excitation by a localized vibrator in the boundary
  layer.
\newblock \textit{J. Appl. Mech. Tech. Phys.} 25:867--873

\bibitem[{Zuccher(2002)}]{ZuccherT}
Zuccher S. 2002.
\newblock \textit{Receptivity and control of flow instabilities in a boundary
  layer}.
\newblock Ph.D. thesis, Politecnico di Milano, Italy

\bibitem[{Zuccher \& Luchini(2013)}]{Zuccher3}
Zuccher S, Luchini P. 2013.
\newblock Boundary-layer receptivity to external disturbances using multiple
  scales.
\newblock \textit{Meccanica} {i}n press

\bibitem[{Zuccher et~al.(2004)Zuccher, Luchini \& Bottaro}]{Zuccher1}
Zuccher S, Luchini P, Bottaro A. 2004.
\newblock Algebraic growth in a {Blasius} boundary layer: optimal and robust
  control by mean suction in the nonlinear regime.
\newblock \textit{J. Fluid Mech.} 513:135--160

\end{thebibliography}

\end{document}